\documentclass[manuscript]{aastex}

\usepackage{epsfig}                   
\usepackage{multirow}

\newcommand{\ctan}{$^{13}$C($\alpha$,n)$^{16}$O~}
\newcommand{\nean}{$^{22}$Ne($\alpha$,n)$^{25}$Mg~}

\shorttitle{s process in Globular Clusters}
\shortauthors{Straniero et al. }

\begin{document}
\title{Heavy elements in Globular Clusters: the role of AGB stars.}

\author{O. Straniero}
     \affil{INAF-Osservatorio Astronomico di Collurania, 64100 Teramo, Italy and
     INFN-sezione di Napoli, 80126 Napoli, Italy}
      \email{straniero@oa-teramo.inaf.it}
\author{S. Cristallo}
     \affil{INAF-Osservatorio Astronomico di Collurania, 64100 Teramo, Italy and
     INFN-sezione di Napoli, 80126 Napoli, Italy}
\author{L. Piersanti}
     \affil{INAF-Osservatorio Astronomico di Collurania, 64100 Teramo, Italy and
     INFN-sezione di Napoli, 80126 Napoli, Italy}

\date{\today}

\begin{abstract}
Recent observations of heavy elements in Globular Clusters reveal intriguing deviations 
from the standard paradigm of the early galactic nucleosynthesis. If the r-process 
contamination is a common feature of halo stars, s-process enhancements are found in a 
few Globular Clusters only. 
We show that the combined pollution of AGB stars with mass ranging between 3 to 6 M$_\odot$
may account for most of the features of the s-process overabundance in M4 and 
M22. In these stars, the s process is a mixture of two different neutron-capture 
nucleosynthesis episodes. The first is due to the \ctan reaction and takes place
during the interpulse periods. The second is due to the \nean reaction and
takes place in the convective zones 
generated by thermal pulses. The production of the heaviest s elements (from Ba to Pb) 
requires the first neutron burst, while the second produces large overabundances 
of light s (Sr, Y, Zr). The first mainly operates in the less-massive AGB stars, 
while the second dominates in the more-massive.
From the heavy-s/light-s ratio, we derive that the pollution phase should last for  
$150\pm 50$ Myr, a period short enough compared to the  
formation timescale of the Globular Cluster system, but long enough to explain why 
the s-process pollution is observed in a few cases only. 
With few exceptions, our theoretical prediction provides a reasonable reproduction of the  
observed s-process abundances, from Sr to Hf. 
However, Ce is probably underproduced by our models,  
while Rb and Pb are overproduced. Possible solutions are discussed.   
\end{abstract}

\maketitle

\section{Introduction}
All the elements heavier than iron are mainly produced by neutron captures\footnote
{A few isotopes are actually synthesized by the so-called p process whose overall
contribution to the elemental abundances is, however, rather small.}. 
There exist two different nucleosynthesis processes of this type,
the slow (s) process and the rapid (r) process \citep{b2hf}.  
Since the typical neutron density of the r process is more than 10 orders of 
magnitude larger than that of the s process, 
significantly different physical conditions are implied and, in turn, 
very different astrophysical environments.
The r process is commonly associated with massive stars. Two are
the proposed scenarios: core-collapse supernovae (type II, Ib and Ic) 
and Neutron-Star mergers. Although none of the proposed astrophysical sites 
has been confirmed by direct observations, the yields of the r process are
commonly found in all the Galactic components, 
very-metal-poor stars included. Such a prompt pollution, demonstrates that
the r process takes place in stars that evolves on a very short timescale
\citep[see][]{sneden2008}.

On the contrary, our knowledge of the s-process site has been greatly 
improved in the last 20 years \citep[for a review see][]{busso1999}.
First of all, it should be reminded that the most abundant products of 
the s process are the so-called {\it neutron-magic nuclei}, whose  
neutron-capture cross section is particularly low compared 
to the cross section of nearby nuclei.  
When the s-process flow encounters a neutron-magic nucleus, 
it acts as a bottleneck, so that its abundance
is greatly enhanced with respect to the nearby non-magic nuclei, for which 
a nearly local equilibrium is established, as given by: $\sigma_A N_A=\sigma_{A-1} N_{A-1}$
\footnote{
$\sigma_A=\frac{< \sigma v>}{v_{th}}=\frac{2}{\sqrt{KT}}\int_{0}^{\infty}E\sigma_{n}(E)exp\left ( -\frac{E}{KT} \right )dE$
is the Maxwellian averaged cross section (MACS) and $N_A$ is the fraction of
isotopes with atomic mass A.}. The most 
important neutron-magic nuclei encountered by the s process are
$^{88}$Sr, $^{89}$Y, $^{90}$Zr, $^{138}$Ba, $^{139}$La, $^{140}$Ce, $^{141}$Pr,
$^{142}$Nd, $^{208}$Pb and $^{209}$Bi. Each of these nuclei corresponds 
to a peak in the distribution of the solar system abundances. 
The first three are the major contributors to the light-s peak, while those 
from $^{138}$Ba to $^{142}$Nd contribute to the heavy-s peak. 

As pointed out in the seminal 
paper of  \citet{b2hf}, the s process follows simple general rules.
Three are the main players: neutrons (or neutron sources), 
seeds (Fe nuclei) and neutron poisons. The latter are light elements that compete 
with the seeds in the neutron-capture nucleosynthesis. 
In this context, a fundamental quantity that characterizes the s process  
is the neutron-to-seed ratio, i.e., $f=\frac{neutrons-poisons}{seeds}$, 
where {\it neutrons, poisons and seeds} represent fractions by number.
As firstly shown by \citet{cameron1957}, the synthesis of the heaviest elements,
such as Pb,
requires a relatively large value of this ratio ($f>20$), while for low values,
namely $f\sim 1$, only light-s are produced. Note that  
the number of seeds directly scales with the metallicity, 
so that the production of the heaviest s elements is generally
favored at low Z \citep{busso1999,cristallo2009}.
As a matter of fact, the cosmic  concentration of lead is the result of the 
pollution caused by low-metallicity AGB stars \citep[e.g.,][]{travaglio2001}. 
Other important quantities that characterize the s process are
the neutron density ($N_n$), the temperature (T) and the timescale (i.e., the 
duration of the s-process episode).
They determine the time integrated neutron flux, or neutron exposure, 
namely $\tau = \int N_nv_{th}dt$\footnote{$v_{th}$ is the thermal velocity,
which depends on T}. Note that the larger the neutron exposure
the larger the probability to overshoot the neutron-magic nuclei.  
Moreover, the Maxwellian averaged cross sections depend on the temperature, while 
the neutron density is important for the various branchings occurring along the
s-process path. Indeed, when a neutron capture produces an unstable nucleus,  
the $\beta$ decay may compete with a further neutron capture \citep[see][]{kappeler1989}. 
For each branching, it exists a 
critical value of the neutron density given by the ratio of the decay rate 
and the neutron-capture rate. When the neutron density is much larger 
than this critical value, the neutron capture is favored with respect to 
the $\beta$ decay, while the opposite occurs 
at low neutron density. In this way, the neutron density determines 
the abundances of the isotopes on the alternative paths.
Some examples are the branchings at $^{79}$Se, $^{85}$Kr, $^{95}$Zr, 
$^{134}$Cs and $^{151}$Sm. 
In general, those isotopes/elements whose production is
sensitive to the neutron density are good estimators of the physical conditions
of the s-process site \citep{lambert1995,abia2001,aoky2003,barzyk2007,vanraai2012,lugaro2014}.

By analyzing the heavy element composition of the solar system, three different 
components of the s process have been formerly identified, 
namely the weak, the main and the strong 
\citep{seeger1965,clayton1967}. Each s-process component implies 
a specific range of neutron exposures and, in turn, a specific range 
of the quantities characterizing different s-process sites,
i.e., $f$, $N_n$, T and the timescale.
  
The weak component, which includes nuclei with $29<Z<40$, 
is synthesized in the He-burning core and, later on, in the C-burning shell of massive stars
\citep[M$> 10$ M$_\odot$,][]{raiteri1991, raiteri1991b, kappeler1994, pignatari2006}. 
The neutron density may vary from $\sim 10^6$ neutrons/cm$^3$, 
in the case of the He burning, up to  $\sim 10^{11}$, for the C-burning. 
Temperatures and timescales are also very different, but the
neutron exposure is similar, namely $\sim0.06$  mbarn$^{-1}$.  
In both cases, neutrons are provided by the \nean reaction, so that  
$^{22}$Ne is a necessary ingredient for the weak-s process.  
In practice, $^{22}$Ne is synthesized during He burning, through the 
$^{14}$N$(\alpha,\gamma)^{18}$F$(\beta^+)^{18}$O$(\alpha,\gamma)^{22}$Ne chain,
where $^{14}$N is that left by the former
CNO burning. Therefore, the fraction (by number) of $^{22}$Ne nuclei available for the 
s process in massive stars is approximately equal to the original fraction of 
C+N+O nuclei. Such an occurrence implies
that the synthesis of the weak component is less efficient at low Z, 
because of the paucity of C+N+O and, in turn, of $^{22}$Ne. 
For instance, during core-He burning, the main neutron poison is $^{25}$Mg that is secondary
like, since it is directly produced by the \nean reaction. As a result, 
the weak process yields decrease roughly linearly with metallicity.
Instead, in the C-burning shell there are
primary like neutron poisons (e.g., $^{16}$O, $^{23}$Na, $^{24}$Mg), 
which do not depend on the metallicity. Therefore, the s-process efficiency in the 
C-burning shell is strongly suppressed at low Z \citep{pignatari2007}.  
 
Recently, \citet{pignatari2008} show 
that in very low-metallicity fast-rotating massive stars, fresh C synthesized by the 3$\alpha$ 
reaction may be transported by meridional circulation into the H-rich envelope, 
thus increasing the amount of C+N+O. They find that this phenomenon would allows
an efficient s-process nucleosynthesis, up to Pb. However, 
in a more recent paper, \citet{frisch2012} argue that this result
is due to the use of a particularly low rate of the  
$^{17}$O($\alpha,\gamma)^{21}$Ne reaction, i.e., that suggested by \citet{desc1993}, 
which is up to a factor of 1000
lower than the values reported in the widely used reaction rate compilations 
\citep{CF88, NACRE}. In the He-burning core of fast rotating massive stars, 
this reaction is expected to destroy most of the 
$^{17}$O released by the poisoning reaction $^{16}$O(n,$\gamma)^{17}$O.
The suppression of the $^{17}$O($\alpha,\gamma)^{21}$Ne 
favors the competitive channel $^{17}$O($\alpha,$n)$^{20}$Ne, so that 
the neutrons subtracted by the $^{16}$O would be recycled.
However, new experiments reinvestigated both channels of the $^{17}$O$+\alpha$, 
confirming previous findings \citep{best2011, best2013}. In particular,  
they find that the $\gamma$ channel is strong enough to compete with the
neutron channel, thus leading to a less efficient neutron recycling.
Fast-rotating massive stars might 
still play a role in the production of the weak component (up to Sr), 
but no significant s-process contribution to heavier elements are expected
\citep[see Figure 14 in][]{best2013}.     

The main and the strong components, which include nuclei with $37<Z<84$, 
are produced by low-mass stars ($1.5<$M/M$_\odot<2.5$)
\citep{straniero1995, gallino1998, cristallo2009, cristallo2011}.
In these stars, recursive thermonuclear runaways of the 
shell-He burning, called thermal pulses (TPs), take place during the AGB phase.
Two important events are connected to the occurrence of these thermal pulses.
First of all, owing to the excess of nuclear energy released by the thermonuclear runaway, 
an extended convective instability takes place within the He-rich layer.
Later on, owing to the expansion powered by He burning, the shell-H burning
dies down and the inner border of the convective envelope can attain the
He-rich zone (third dredge up - TDU). 
The s-process nucleosynthesis in low-mass stars mostly occurs 
during the relatively long interpulse
period ($\sim 10^5$ yr), namely the time elapsed between two subsequent 
thermal pulses, in a thin radiative layer located at the top of 
the He-rich zone \citep{straniero1995}. This layer is known as the $^{13}$C pocket, because
it is enriched in $^{13}$C. The neutron source is the \ctan reaction, which requires  
a temperature of $\sim90-100$ MK and releases low-density neutron fluxes,
i.e., about $10^7$ neutrons/cm$^{3}$, and neutron exposures between 0.1 and 0.4 
mbarn$^{-1}$ \citep{gallino1998}.
A second neutron burst giving rise to a higher neutron density 
($>10^{11}$ neutrons/cm$^{3}$) is due to the marginal activation of the \nean reaction 
within the convective zone generated by a thermal pulse, where the temperature 
may exceed 300 MK. In this case, the timescale is rather short ($\sim 1$ yr), 
so that the resulting neutron exposure is lower than that
of the first neutron burst.
These low-mass stars are the main contributors to the s-process elements 
in the solar system. However, because of their long lifetime ($\ge1$ Gyr), 
it appears that they cannot have contaminated the gas from which the galactic halo formed. 

This is the standard paradigm for the heavy element composition of the halo.
In practice, only r-process yields are expected in fossil records of the early Galaxy, 
the s process
being hampered by the secondary nature of the neutron sources in massive stars (weak component) 
and by the too long lifetime of low-mass AGBs (main and strong components).   
Spectroscopic studies generally confirm such a scenario: single halo stars are 
r-process enriched, but s-process poor \citep[see][and references therein]{sneden2008}. 
Exceptions are the CEMP-s (Carbon-Enhanced-Metal-Poor stars, 
where the ``s'' stay for  s-rich). In this case however, 
the s and the C enrichments are a consequence of 
mass transfer or wind accretion in binary systems, a process occurring on a longer 
timescale \citep[see][and references therein]{bisterzo2012, lugaro2012}.  

In this context, recent spectroscopic studies of Globular Clusters (GCs)
revealed a rather different scenario. While the r-process yields generally 
appear similar to those observed in halo field stars,
some GCs show a clear signature of the s-process main component pollution.
The few GC stellar populations where an s-process enrichment has been 
discovered are: M4 \citep{yong2008, dorazi2013a},  
$\omega$-Cen, only stars with [Fe/H]$>$-1.6 \citep{smith2000, johnson2010, dorazi2011},
and the redder main sequences of M22 \citep{roederer2011} and NGC1851 \citep{gratton1851}.
Recently, s-process overabundances have been also found in M2 stars \citep{lardo2013}. 
Other clusters, like M5 \citep{ivans2001, yong2008}, as well as the most metal-poor 
stellar populations of $\omega$-$Cen$, M22 and NGC1851, present
a ``normal'' halo distribution of the heavy elements characterized by a 
pure r-process pollution.  
These challenging observations represent a further
evidence of the existence of multiple stellar populations in GCs. Nevertheless, 
at variance with
other spectroscopic anomalies, 
such as the O-Na anticorrelation
\citep[][and references therein]{gratton2012}, the s-process enhancement 
is not a common feature of 
the majority of the GCs in the Milky Way. Therefore, 
a different class of polluters should be responsible for the 
heavy-element anomalies. 
Such a conclusion is also supported by the fact that
in M22 and NGC1851 the O-Na anticorrelation 
is observed in both s-rich and s-poor stars of the 
same cluster.
Moreover, all stars in M4 show a similar overabundance of the s elements 
but this enrichment is uncorrelated with the [Na/Fe].
More intriguing, some spectroscopic indexes, which depend on the metallicity
of the polluters, do not match the theoretical expectations 
for low-mass AGB stars, which are considered the most important producers
of the galactic s-process main and strong components.  
In particular, the ratio between heavy-s 
(Ba, La or Nd) and light-s elements (Sr, Y or Zr) are found in solar 
proportions ([hs/ls]$\sim0$]), while an excess of heavy-s is expected at low Z. 
Therefore, the polluters responsible for such a peculiar chemical 
pattern cannot be the same stars responsible for the bulk of the s-process 
yields in the Galaxy.

In this paper we study the characteristics of the 
s-process nucleosynthesis in metal-poor AGB stars of low and intermediate mass. 
We will discuss, in particular, 
the variations of the nucleosynthesis outcomes with the stellar mass.  
In the next section we review the most important inputs physics and how they are
included in our stellar evolution code. In section 3  
we analyse the operation of the two neutron sources active in thermally pulsing AGB
stars. This analysis is based on the models presented in section 4. 
The theoretical yields we derive from these models may be used 
to test various scenarios for GC formation that 
have been proposed to explain photometric and spectroscopic evidences 
of multiple populations, among which: multiple photometric sequences,
star-to-star variations of the chemical composition, which cannot be ascribed to 
internal physical processes, and anomalous color dispersion  
of horizontal branch stars or the so-called second parameter problem 
\citep[for a recent review see][]{gratton2012}.     
Several hypotheses about the GC formation have been proposed to explain the new 
observational framework, such as:
inhomogeneities of the primordial material, merging of smaller stellar systems,
pollution with external material felt into the gravitational potential well 
of the cluster and various self-pollution scenarios. 
Which of these scenarios can also provide an explanation for the s-process enhancements
observed in a few GC stellar populations?  Which stars are responsible for the s-process contamination 
in GCs? What are the special conditions determining 
the onset of this peculiarity? These issues are addressed in section 5 and 6.
We show, in particular, that AGB stars with mass ranging between 
3 to 6 M$_\odot$ can produce the yields necessary to reproduce the observed 
heavy-element anomalies. In this case, we find that the time elapsed between 
the formation of the polluters and that of the polluted stellar populations  
should be of the order of 100-200 Myr.              

\section{The stellar evolutionary code}\label{codedescription}
All the stellar models presented in this paper have been computed by means of
our FUNS code (FUll Network Stellar evolution code)\footnote{Such a code has been 
derived from the FRANEC code \citep{cs1989, cls1998}.}. 
As illustrated in \citet{straniero2006} \citep[see also][]{cristallo2009, cristallo2011},
it includes a full nuclear network of 
about 500 isotopes (from $^1$H to $^{209}$Bi) and more than 1000 nuclear reactions,
coupled to the standard set of 1d hydrostatic differential equations 
that describe the physical structure. Rotation has been
recently included and the resulting effects on the 
s-process nucleosynthesis occurring in low-mass AGB stars have been discussed in 
\citet{piersanti2013}. The models here presented are for non-rotating stars.
Rotating models for intermediate mass AGB stars will be presented in a forthcoming paper. 

The occurrence of recursive thermonuclear runaways makes the computation 
of AGB evolutionary sequences 
and the related nucleosynthesis a challenging task for stellar modelers.  
Numerical algorithms and input physics should be particularly accurate to properly follow 
significant variations of the physical and chemical structure on relatively 
small temporal and spatial steps. Many efforts have been made to improve the physical 
description of these stars and a qualitative agreement is generally found between models 
produced by different groups, although quantitative results may be rather different.
In this section we review the physical processes expected to produce major uncertainties 
on AGB calculations and how they are treated in the FUNS code.

\subsection{Mass loss}
The AGB mass-loss rates are usually estimated from infrared colors or CO rotational lines.
Thanks to the recent progress of the infrared astronomy, 
our knowledge of the AGB mass loss has been significantly improved \citep[see, e.g.,][and references therein]{groenewegen2009}. 
Nonetheless, a general prescription to be used in stellar model calculations 
is far from being definitely established. 

AGB stars are long period variables, Miras or irregulars. 
In this context, the mass-loss rate versus period relation 
is an appealing tool for the purpose of stellar model calculations \citep{vw1993, whitelock1994,
schoier2001, whitelock2003, winters2003, groenewegen2009}.
 \citet{vw1993} firstly use a  mass-loss rate versus period 
relation to calculate AGB models of different masses. In \citet{straniero2006}, 
we update this relation by means of a more extended set of 
infrared data. In general, these measurements show that the mass-loss rate remains quite 
moderate, namely $10^{-8}<dM/dt<10^{-7}$, for $\log P(days)<2.5$. For larger periods the mass-loss 
rate steeply increases and attains maximum values for $\log P>3$. This 
upper limit (a few $10^{-5}$ M$_\odot$/yr) coincides with the superwind phase, 
which is  dominated by a radiation-pressure driven wind \citep{vw1993}. 
It should be noted, however, that the observed spread of the mass-loss rate versus period 
relation is rather large. For a given period, the mass-loss rate 
may vary up to a factor of 10 \citep[see, e.g.,][Figure 5]{straniero2006}. 
Nevertheless, it appears that 
the mass loss versus period relation is a universal feature of AGB stars, 
independent of stellar mass and envelope composition. For instance,
\citet{whitelock2003} do not find differences between O-rich and C-rich stars, while
\citet{groenewegen2009} exclude a variation with the metallicity. 
Therefore, it appears that the same mass-loss rate versus period relations can be applied to 
AGB models of different mass and initial composition. 
On the contrary, other mass-loss prescriptions, 
such as the empirical formula derived by \citet{vanloon2005} or the semi-empirical 
relation provided by \citet{blocker1995}, apply to models within a more restricted 
range of stellar parameters (i.e., initial mass and composition). 
It should be remarked that most of the 
available mass-loss studies are based on stars of the Galactic 
Disk, Galactic Bulge and Magellanic Clouds. Little is known about AGB 
mass loss for stars with metallicity as low as that of the Galactic GCs. 

The mass-loss rate versus period formula used to compute all the AGB models presented in 
this paper is that described in \citet{straniero2006}. For the pre-AGB phase a classical 
Reimers' mass-loss rate ($\eta=0.4$) is used.

\subsection{Super-adiabatic convection}
\begin{figure}
\includegraphics[angle=0,width=\columnwidth]{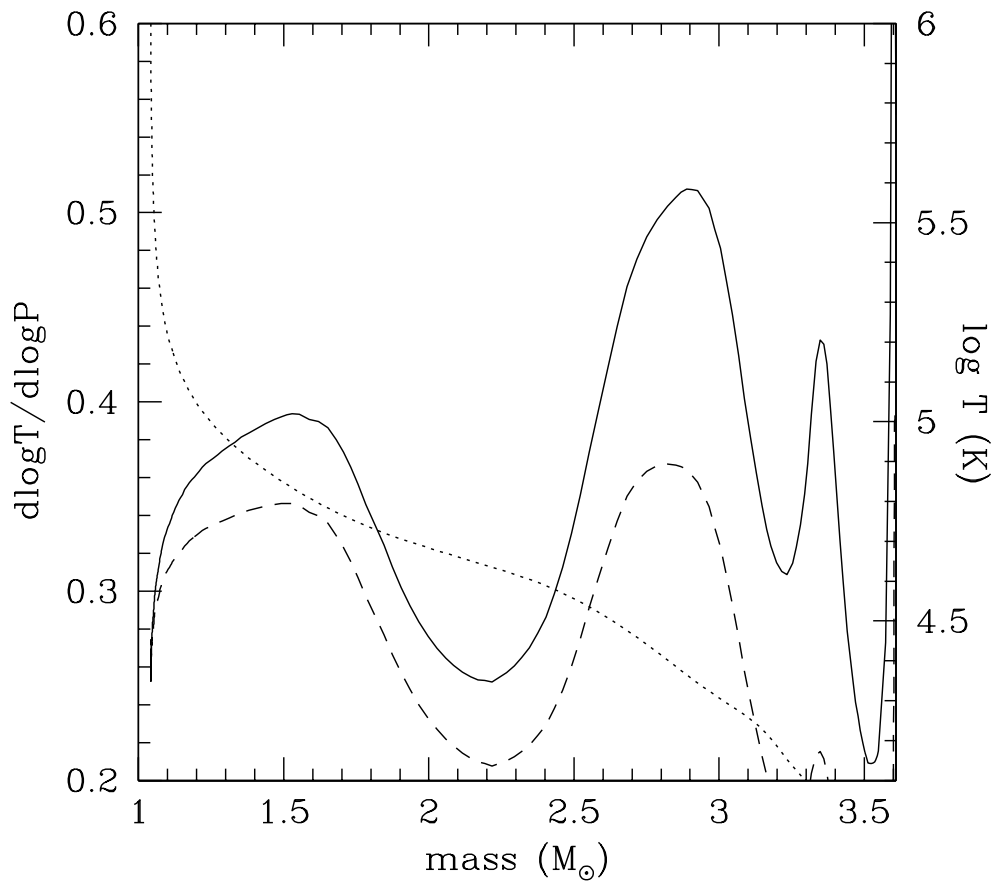}
\caption{Temperature gradients within the convective envelope of a 6 M$_\odot$ 
model during the 50th interpulse period: effective gradient (solid line), 
adiabatic gradient (short-dashed line).
The temperature profile is also shown (long-dashed line). Note that the 
temperature at the 
convective boundaries, i.e., 85 MK (internal) and  4000 K (external), are out 
of the Y axis range.}\label{grad}
\end{figure}

In the external convective layers of red giant stars, convective heat transfer only partially
accounts for the whole outgoing energy flux. In this case, the effective 
temperature gradient is larger than the adiabatic one. This is particularly important 
for AGB stars, where more than 90\% of the mass of the convective envelope undergoes  
super-adiabatic conditions (see Figure \ref{grad}). 
The mixing-length theory \citep[MLT;][]{bv1958} is widely used 
to evaluate the super-adiabatic temperature gradient. 
Being a phenomenological theory, it implies a number of free parameters, 
usually reduced to just one called  $\alpha$, i.e., the ratio of the average mixing length 
to the pressure scale height. 
Note that there exist different versions of the MLT, so that the physical 
meaning of $\alpha$ may differ from author to author. 
In our calculation we adopt the formalism described by \citet{coxgiuli1968}. 
An alternative phenomenological approach is that proposed by  
\citet{cm1991} (hereinafter CM).  They consider the full turbulent energy 
spectrum and set the convective scale length equal to the geometrical depth 
from the top of the convective region. Comparisons between MLT 
and CM models show that the CM formalism 
cannot be reproduced by the MLT with any constant value of $\alpha$ 
\citep{mazzitelli1995}. In practice, the mixing-length free parameter 
is usually calibrated by reproducing the solar radius with a 
standard solar model. However, there are no physical reasons to believe that a unique value of 
$\alpha$ is suitable for any stellar model. Therefore, it may be possible that 
$\alpha$ should be varied along an evolutionary sequence. This issue has been
investigated by \citet{freytag1999} by means of multidimensional
radiation hydrodynamics (RHD) simulations covering the range of effective 
temperatures, gravities, and compositions typical of MS and RGB stars of 
Galactic GCs.  
They found that RGB models computed with the $\alpha$ parameter derived 
from RHD are slightly cooler than those computed adopting a solar 
calibrated $\alpha$, (less than 10\%), while the CM models predict too low 
effective temperature. In practice, RGB models would require an $\alpha$ value
larger than that needed to reproduce the solar radius.
A direct comparison with measured effective temperatures 
of RGB stars have been also reported by \citet{ferraro2006} 
\citep[see also][]{chieffi1995}. This study confirms previous findings of \citet{freytag1999}.   
Note that a similar comparison cannot be easily obtained with AGB stars, because of 
the recursive variations of the effective temperature caused 
by thermal pulses. Since the convective envelope of an AGB star is more similar
to that of a RGB star than to the solar convective envelope,  
all the models presented in this paper have been computed with the MLT 
and $\alpha$ calibrated on 
GC RGB stars, namely $\alpha=2.1$, as reported in \citet{ferraro2006}. 

In low-mass AGB stars, a variation of $\alpha$ affects the efficiency of the TDU
\citep[e.g.,][]{cristallo2009,cristallo2011}. In addition, 
since the effective temperature depends on $\alpha$, the mass-loss rate is
also changed. Both these effects modify 
the heavy element yields. Nevertheless, the relative abundances of Pb, 
heavy-s and light-s are marginally affected.
In massive AGB, a variation of $\alpha$ also affects the maximum temperature attained at 
the bottom of the convective envelope and, in turn, the  
nucleosynthesis in the convective envelope and 
the stellar luminosity \citep[e.g.,][]{IR1983,ventura2005}. 

\subsection{Time dependent convection}
Owing to the complex nucleosynthesis taking place within the convective regions of AGB stars,
which involve a great number of nuclear species, 
some nuclear burning timescales may be comparable or smaller than the convective turnover timescale.
In that case, the assumption of instantaneous mixing, which is usually adopted in computations 
of pre-AGB stellar models, is no more valid. 
In case of stars undergoing hot bottom burning \citep[HBB, see][]{IR1983}, 
high temperatures are attained in the convective envelope during the interpulse period. 
As an example, for the model shown in Figure \ref{grad}, the temperature at the 
bottom of the convective envelope is about 85 MK, which is large enough for the activation of 
the CN cycle.    
Note that the extension of the convective envelope of these giant stars 
is of a few hundreds R$_\odot$ and the average convective velocity, 
as derived by means of the MLT, 
is of a few $10^5$ cm/s, so that the convective turnover timescale is of the order 
of $10^8$ s. Then, it can be easily verified that the burning timescales of several 
isotopes involved in the HBB nucleosynthesis are shorter than the convective turnover timescale.
For instance, the $^{12}$C burning timescale becomes shorter than $10^8$ s for T$>60$ 
MK\footnote{The $^{12}$C burning timescale is given by $\tau_{12}=\frac{1}{\rho r_{12,1}X_H}$,
where $r_{12,1}$ is the rate of the $^{12}$C$(p,\gamma)^{13}$N reaction, $\rho$ is the density and $X_H$
is the hydrogen mass fraction.}. Therefore, nuclear reactions are faster than the 
convective mixing, so that partial mixing takes place. 
Note that the effects of the HBB are overestimated 
when instantaneous mixing is assumed, with important consequences on 
both the nucleosynthesis and the HBB contribution to the luminosity.
Concerning the convective zone generated by a thermal pulse, convection is usually faster than
the $\alpha$ captures, so that the majority of the
nuclear species are efficiently mixed. This is not the case of neutrons and protons. 
The neutron burning timescale is extremely short, so that the neutrons released 
by the \nean reactions are
suddenly captured as if they were in a radiative environment. On the other hand,
within a C-rich environment where the temperature is of the order of 200-300 MK, as in the 
convective zone generated during a TP,
the proton mobility is also very limited, mainly because of the $^{12}$C$(p,\gamma)^{13}$N reactions.
Also in this case deviations from the instantaneous mixing must be properly accounted.  

The time dependent mixing scheme we use has been described in \citet{straniero2006}. 
It has been derived from an algorithm originally proposed by \citet{se1980}.
In brief, the degree of mixing is calculated by means of the following relation:
\begin{equation}
X_j=X_j^o+\frac{1}{M_{conv}}{\sum}_{k}(X^o_k-X^o_j)f_{j,k}{\Delta}M_k
\end{equation}
where $X_j^o$ and $X_j$ are the mass fractions in the mesh-point $j$ 
at time $t$ and $t+\Delta t$, respectively. The summation is extended over 
the whole convective zone. ${\Delta}M_k$ is the mass of the mesh-point 
$k$, while $M_{conv}$ is the total mass of the convective zone.
The damping factor $f_{j,k}$ is:
\begin{equation}
f_{j,k}=\frac{{\Delta}t}{{\tau}_{j,k}}    
\end{equation}
if ${\Delta}t<{\tau}_{j,k}$,  or
\begin{equation}
f_{j,k}=1                  
\end{equation}  
if ${\Delta}t\ge{\tau}_{j,k}$.
Here ${\Delta}t$ is the time step and $\tau_{j,k}$ is the mixing turnover time
between the mesh-points {\it j} and {\it k}, namely: 
\begin{equation}
\tau_{j,k}=\int_{r(j)}^{r(k)} \frac{dr}{v(r)}={\sum}_{i=j,k}\frac{{\Delta}r_i}{v_i}   
\end{equation}
The mixing velocity ($v_i$) is computed according to the MLT.     

\subsection{Instability of the convective border, third dredge up and $^{13}$C pocket}
When the convective envelope penetrates the H-exhausted core, a steep variation of the 
composition takes place at the convective boundary: the H mass fraction drops 
from about 70\%, within the fully convective envelope, 
down to zero, in the underlying radiative core. The composition gradient induces 
a sharp variation of the radiative opacity and, in turn, an abrupt change of the 
radiative temperature gradient. In these conditions, 
the precise location of the convective border, as defined by the neutrality 
condition $\nabla_{rad}=\nabla_{ad}$\footnote{
$\nabla_{rad}=\left (\frac{\partial \log T}{\partial \log P}  \right )_{rad}$ and
$\nabla_{ad}=\left (\frac{\partial \log T}{\partial \log P}  \right )_{ad}$ are 
the radiative and the adiabatic temperature gradients, respectively.},
becomes highly uncertain. Indeed, 
even a small perturbation 
causing mixing across the boundary layer induces an increase of $\nabla_{rad}$
in the radiative stable zone so that the convective instability moves toward 
the interior. This situation is commonly encountered in AGB stellar models 
when a bare Schwarzschild's criterion is used to fix the convective boundaries.   
It affects both the second and the third dredge up \citep{becker1979, castellani1990, 
frost1996,marconi1998,mowlavi1999}. 
However, if the effects of such an instability on the second dredge up are probably marginal
\citep{marconi1998},  
this is not the case of the third dredge up. Various attempts have been made to
overcome such a problem, but a satisfactory solution is still lacking.
For instance, \citet{boothroyd1988} extend to the AGB
a method originally developed by \citet{castellani1985}
to treat a similar instability occurring at the outer edge 
of the convective core during the core-He burning phase. 
In practice, they try to extend the convective zone, namely: 
if after mixing the previously stable mesh points become 
unstable, a further extension of the convective zone is applied, otherwise 
mixing is limited to the mesh-points where $\nabla_{rad}\geqslant\nabla_{ad}$.
A different approach has been followed by \citet{frost1996}.
In stellar evolution codes based on the implicit Henyey method, 
the convective boundaries are usually calculated once per time step. 
Instead, \citet{frost1996} recalculate the convective boundaries and the
corresponding new abundances after each Henyey iteration. 
This approach should allow to take into account the  
feedback of the physical structure due to the variation of the chemical composition
induced by mixing. However, severe numerical instabilities are encountered
when this procedure is adopted. To overcome such a problem, 
Frost and Lattanzio set a maximum number of Henyey iterations after which 
the integration is stopped, even if the stellar structure equations are not satisfied.
Note that both \citet{boothroyd1988} and \citet{frost1996} assume, as usually done,  
i) instantaneous mixing within the convective
zone and no extra-mixing beyond the convective boundaries.    
Actually, the transition between the fully-radiative and the fully-convective 
zone most likely occurs over a somewhat extended layer, 
where only a partial mixing takes place. 
In general, hydrodynamical models of stellar convection 
confirm the existence of this transition layer 
\citep[e.g.,][and reference therein]{freytag1996,canuto1998,canuto1999,young2003}. 
In particular, \citet{freytag1996} 
find that the average convective velocity should drop exponentially below 
the shallow fully-convective envelope of A-type stars and cool white dwarfs.
Unfortunately, a limited number of hydrodynamical investigations 
have been carried out so far for AGB stars undergoing TDU \citep[see, e.g.,][]{young2003},
so that generalized prescriptions for the  
extension of the transition layer and the strength of the decline 
of the average convective velocity are not available yet for AGB computations. 
Nevertheless, owing to the relevance of this phenomenon,
AGB models obtained by assuming an exponential decline at the convective boundaries
have been developed by various authors 
\citep{herwig1997,herwig2000,mowlavi1999,chieffi2001,straniero2006,cristallo2009,cristallo2011}. 
In these models, a smooth and stable variable H profile forms at the inner border 
of the convective envelope, so that the instability occurring at the third 
dredge up is removed. 
In addition, the presence of this H profile left in the zone highly 
enriched in He and C provides the conditions for the formation of the 
$^{13}$C pocket.
Note that such an approach necessarily requires the introduction of a free 
parameter, i.e., the strength of the exponential decline 
(hereinafter, the $\beta$ parameter). 
For instance, in our computation the average convective velocity
within the convective-radiative transition layer is given by:
\begin{equation} \label{param}
v=v_{0}\exp{\left(-\frac{d}{\beta H_P}\right)}
\end{equation}
\noindent
where {\it d} is the distance from the formal convective boundary,
$v_{0}$ is the average velocity of the last unstable mesh point,
$H_P$ is the pressure scale height and $\beta$ is a free parameter.
In principle, each convective boundary would require a different $\beta$ value. 
Note, however, that within the fully convective zones we calculate $v$ by means 
of the MLT. 
In this case $v\propto(\nabla_{rad}-\nabla_{ad})$, so that $v_{0}=0$
when, as usual, the neutrality condition is fulfilled at the convective boundary.
In practice, since we consider the average convective velocity, 
convective overshoot is neglected in our models\footnote{With the term
``convective overshoot'' we intend a mixing beyond the convective boundary 
in a layer where $\nabla_{rad}-\nabla_{ad}<0$ both before and after the mixing 
\citep[see, classical papers by][]{bertelli1985,maeder1987}.}. On the other hand,
$v_{0}\gg0$ during a dredge up (because $\nabla_{rad}\gg\nabla_{ad}$),
so that an extramixing naturally araises.
Two are the main consequences of this approach, namely i) 
a deeper dredge up and ii) the development of a $^{13}$C pocket.
The total mass of $^{13}$C in the pocket depends on both the 
strength of the exponential decline ($\beta$ in equation 5) and  the adopted mixing scheme.  
Our tests show that when a diffusive scheme is adopted, as in several extant 
stellar evolution codes, the resulting $^{13}$C mass is too small to allow 
a sizeable s-process nucleosynthesis for any choice of $\beta$  
\citep[see also][]{herwig2000}.
Instead, in our scheme the degree of mixing between two 
mesh-points depends linearly on the inverse of their reciprocal 
distance and on the 
corresponding turnover timescale (see previous section). 
By means of this mixing scheme and with a proper choice of $\beta$,
we are able to obtain 
sufficiently large $^{13}$C pockets to account for the bulk of the s-process 
enhancements measured in AGB stars of relatively low mass (1.5-2.5 M$_\odot$).  
The variation of the $^{13}$C pocket with $\beta$ and its calibration have been extensively 
discussed in \citet{cristallo2009}.

\subsection{Radiative opacity and equation of state}
The occurrence of recursive dredge-up episodes produces significant 
changes in the chemical composition of the stellar envelope. 
In principle, a stellar evolution code should account for this phenomenon.  
In practice, only variations of the main constituents are usually 
considered. In particular, extant stellar models are 
based on radiative opacity tables for fixed composition (i.e., scaled solar) 
except for H and He. This procedure substantially underestimates the 
radiative opacity of the cool atmosphere of evolved AGB stars, 
which are highly enriched in C and N.
Besides the local thermodynamic conditions, the concentration of the 
various molecular species basically depends on the atomic abundances. 
In this respect, an important quantity is the C/O ratio.  
Among the various molecular species involving C atoms, CO has the larger 
dissociation energy, so that for C/O$<$1 almost all the C atoms participate 
to the formation of this molecule, while the oxygen atoms in excess are free 
to form other molecules, such as TiO and H$_2$O. However, when as a consequence 
of the TDU C/O becomes larger than 1, carbon-bearing molecules, 
e.g., C$_2$, CN, C$_2$H$_2$, and C$_3$, dominate the radiative opacity.
In addition, if the bottom of the convective envelope attains temperature
that are large enough for the activation of the CN cycle, some of the C dredged 
up is converted into N.  
\citet{marigo2002} made a first step toward a correct description of the 
abundance changes in the calculation of opacity coefficients by estimating 
molecular concentrations through dissociation equilibrium calculations. 
Her main finding is the substantial decrease of the effective temperature
of C-star models, which implies a huge increase of the mass-loss rate. 
More recently, \citet{cristallo2007} \citep[see also][]{marigo2009}
presented new opacity tables with variable amount of C and N. 
The effects of the adoption of these new opacity tables are particularly 
strong at low metallicity for which the C+N 
enhancement in the envelope may be as large as a factor 1000.
  
The model presented in this paper have been computed with the following 
prescriptions for the radiative opacity. For T$\le 10^4$ K we use   
tables that allow arbitrary enhancements of both C and N and include 
both atomic and molecular opacity sources \citep{cristallo2007}.
For larger temperature, we have generated specific opacity tables from the OPAL
facility \citep{iglesias1996}. Also in this case arbitrary enhancements 
of C and N are allowed. 

The equation of state (EOS) is another critical ingredient of AGB models.
An EOS suitable for the high-density regime of the stellar core should account for
electron degeneracy and electrostatic interactions \citep[see][]{straniero1988}. In addition, 
relativistic corrections should be considered for density $\sim10^6$ g/cm$^3$,
as it occurs near the stellar center. Partial degeneracy of the electron component 
takes place in the transition layer between the CO core and the He-rich intershell.
At the opposite, in the cool H-rich envelope, partial ionization of atoms and, more outside,
molecular recombination produce sizeable modifications of the relevant thermodynamic quantities,
such as pressure, specific heat or adiabatic gradient.
After various tests devoted to quantify the different contributions to the EOS 
and their implications for the models, we have adopted the following prescriptions.
For T$<10^6$ K we use the EOS2005 tables provided by the OPAL collaboration \citep{rogers1996}.
These temperatures are usually attained in a great portion of the H-rich envelope, where 
partial ionization and molecular recombination of the most important chemical 
species occur. A double interpolation, on hydrogen mass fractions and Z, is performed 
to account for the variations of the chemical composition due to nuclear burning and/or mixing.
Note that in this way the modifications of the envelope composition due to
the TDU are not properly taken into account. Indeed, for a given Z, the
relative abundances of elements heavier than He are fixed to the scaled-solar values.
Nevertheless, we have verified that at variance with the radiative opacity,
the interpolation on the total Z provides a good approximation for the EOS. 
This method is not adequate for 
the H-exhausted core, where the contribution to the EOS of elements heavier than 
He is more important. For this reason, at temperature larger than $10^6$ K we use EOS tables 
for pure elements, as described in \citet{straniero1988} and \citet{prada2002}. 
This EOS implies full ionization, which is a quite good approximation for 
T$>10^6$ K\footnote{Actually, at T$\sim10^6$ K, this approximation is 
very good for hydrogen and helium, almost true for the most abundant metals (C, N and O), 
while iron ions may hold a few of their inner electrons.}. 
Therefore, additivity laws that apply to specific thermodynamic quantities, such as
pressure, volume and all the state functions, are exploited to combine 
the contributions of the various chemical species. 
The transition across the T$=10^6$ K boundary is sufficiently smooth.

\section{The two neutron bursts}\label{nsources}
As recalled in the Introduction, two different neutron sources are active in 
AGB stars undergoing TPs and TDU. 
In this section we illustrate, on the base of stellar models computed with the FUNS code,
the operation of these neutron sources in low metallicity stars of low and intermediate mass. 
The discussion will be focused on the parameters affecting 
the nucleosynthesis.

\subsection{The radiative \ctan neutron source at low Z}\label {13C}
\begin{figure}
\includegraphics[angle=0,width=\columnwidth]{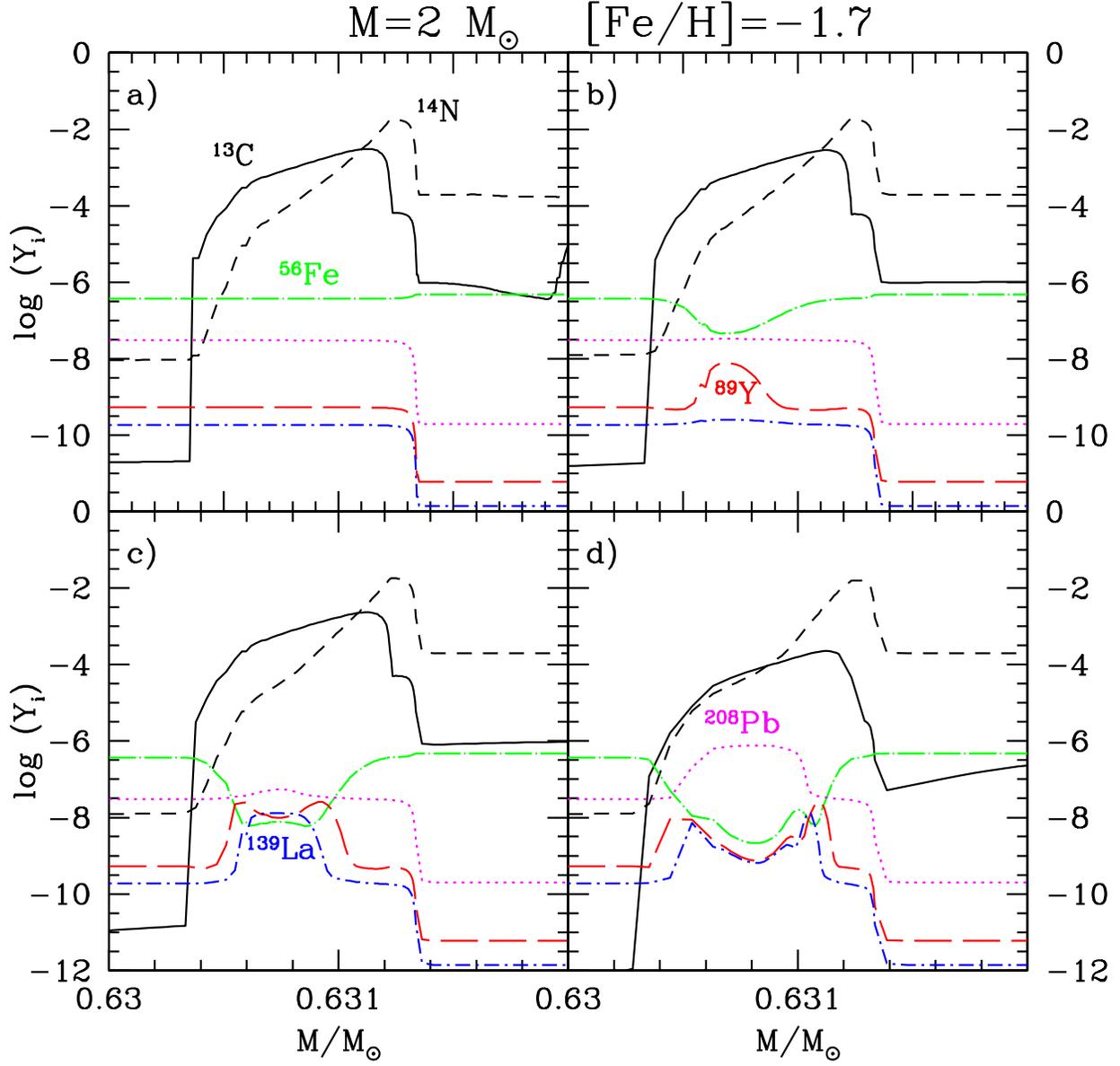}
\caption{Evolution of some chemical species within the third fully developed
$^{13}$C pocket of the 2 M$_\odot$ model. Panel a): the pocket formation is  
completed; panel b): the early phase of the radiative s-process nucleosynthesis,
characterized by the production of light-s elements; panel c): intermediate phase
during which heavy-s elements are mainly produced; panel d) the late part
of the s process, when a huge amount of Pb is synthesized into the pocket.}\label{tasca}
\end{figure}

The TDU is a necessary condition for the formation of the $^{13}$C
pocket. When the convective envelope penetrates the H-exhausted 
core, down to a layer where the mass fraction of the $^{12}$C produced 
during the previous thermal pulse is about 20\%, and, then, recedes,
it leaves a variable H profile (see section 2.4). 
Later on, when this region contracts and heats up, the 
$^{12}$C$(p,\gamma)^{13}$N$(\beta^+)^{13}$C chain starts to produce $^{13}$C. 
Then, a $^{13}$C pocket forms in the innermost tail  of the variable H profile
left by the TDU. More outside, 
owing to the larger H mass fraction, the CN cycle is completed
and $^{14}$N, rather than $^{13}$C, is produced (see panel a) in Figure
\ref{tasca}). The temperature in the zone occupied by the thin $^{13}$C pocket, 
which now contains up to a few $10^{-6}$ M$_\odot$ of $^{13}$C, 
continues to increase during the 
interpulse period. Later on, when the temperature attains $\sim 90$ MK, the neutron-capture 
nucleosynthesys powered by the \ctan reaction begins.
Initially, the iron seeds in the pocket are rapidly consumed to produce 
light-s isotopes, up to the first bottleneck corresponding to the magic nuclei 
$^{88}$Sr, $^{89}$Y and $^{90}$Zr (see panel b) in Figure \ref{tasca}). 
Then, since at that time most of the iron has been already consumed within the pocket,
these light-s nuclei become the main seeds of the s process. 
While these nuclei are consumed, those belonging to the second
bottleneck, such as $^{138}$Ba or $^{139}$La, are accumulated
(panel c) in Figure \ref{tasca}). When the amount of heavy-s 
overcomes that of the light-s, the amount of $^{13}$C 
is still large and also $^{138}$Ba or $^{139}$La become seeds, 
while $^{208}$Pb and $^{209}$Bi are accumulated (panel d) in Figure \ref{tasca}). 
Summarizing, the large
excess of $^{13}$C with respect to the original iron seeds in the pocket leads
to the production of a large amount of lead, the end-point 
of the s process.  
This occurrence is a common feature of low-mass AGB models with 
[Fe/H]$< -1$. By increasing the metallicity,
the Pb produced by the $^{13}$C burning decreases, but the heavy-s are 
overproduced with respect to the light-s \citep[see][]{busso1999}. Finally, if the number of Fe nuclei in the pocket is 
comparable to (or larger than) the number of $^{13}$C nuclei, 
the main s-process yields are light-s. Such a condition is attained for 
Z$\ge$Z$_\odot$. 
Note that the mass extension of the $^{13}$C pocket decreases as the core mass
increases. As a result, the largest pockets are those forming
at the beginning of the thermal pulse AGB phase.
For this reason, in low-mass AGB stars (M$<$ 3 M$_\odot$),
the s-process nucleosynthesis is dominated by the few (3 or 4) initial 
$^{13}$C bursts \citep{cristallo2009}.
For the same reason, the $^{13}$C pockets are smaller in more massive AGB,
so that the s-process contribution of the radiative \ctan neutron-capture nucleosynthesis 
decreases as the stellar mass increases.
In massive AGB stars, additional phenomena 
affecting the development of the $^{13}$C pocket should be considered:
the hot third dredge up \citep[HTDU, see][]{goriely2004}
and the HBB. 
In the first case, when the convective envelope 
penetrates into the H-exhausted core (TDU), it encounters hotter layers.
If the core mass is large enough, the temperature may be sufficiently high to 
activate proton capture reactions. Then, the energy released by the nuclear reactions
contrasts the convective instability that is pushed outward.
This phenomenon has two effects. First, it rises a barrier 
that limits the TDU. Second, when the convective envelope
recedes, it leaves a steeper H profile compared to that obtained in low-mass
stars. Therefore, the resulting $^{13}$C pockets are smaller.  
The HBB takes place during the interpulse period in the more massive AGB and 
in super-AGB stars\footnote{As usual, {\it AGB stars} are those stars that 
enter in the AGB phase just after 
the He-burning phase, while in {\it super-AGB stars} the AGB phase follows the  
C-burning phase. The transition mass between the progenitors of AGB and 
super-AGB stars, the so-called $M_{up}$, is rather uncertain and depends on both 
metallicity and He content \citep{becker1979}.}. 
In this case, the convective envelope attains layers where the H-burning 
nucleosynthesis takes place.
Then, fresh fuel stored in the cooler portion of the envelope  
is continuously brought into the burning zone, so that the 
H-burning rate increases. As firstly demonstrated by \citet{straniero2000}, 
this rate determines i) the physical conditions at the He ignition, ii) the 
power of the consequent thermal pulse and iii) the deepness of the following TDU.
In particular, 
the faster the H burning, the weaker the He flash and, in turn, the shallower the following TDU. 
In summary, the combined action of HTDU and HBB prevents the formation of sizeable 
$^{13}$C pockets and limits the penetration of the convective instability into 
the H-exhausted core, so that the s-process nucleosynthesis due to the \ctan neutron 
burst is suppressed.   
The upper mass limit depends on the metallicity.
At Z$=10^{-4}$, we found that the $^{13}$C neutron source provides a non-negligible 
contribution to the s-process nucleosynthesis for M$\le4$ M$_\odot$, while at 
Z$=$Z$_\odot$ this limit rises up to $\sim 5$ M$_\odot$. In any case, 
the core mass should be lower than about 0.9 M$_\odot$.
Let us note that the precise value of the maximum mass (or core mass) 
for the development of sizeable $^{13}$C pockets depends on the adopted mixing 
scheme (see section 2.3).
However, as far as we know, no other authors have investigated this limit by means 
of self-consistent stellar models. Indeed,
a proper treatment of both HBB and HTDU requires stellar model  
calculations performed with a sufficiently extended nuclear network 
coupled to the stellar structure equations. 
Only in this case, the calculation may account 
for all the energetic feedbacks on the physical structure. 
Nonetheless, the shrinking of the $^{13}$C pockets
in the more massive AGB is a widely accepted phenomenon \citep{goriely2004}.  
For instance, \citet{lugaro2012}, in their post-process calculations, do not include the  
nucleosynthesis associated to the $^{13}$C pocket in models with M$>4.5$ M$_\odot$.    
     
\subsection{The convective \nean neutron source at low Z}\label{22NE}
The second neutron burst occurs under very different environmental conditions.
First of all, the temperature required to activate the \nean reaction is definitely 
larger, namely about 300 MK. The maximum temperature reached at 
the bottom of the convective shell generated by a thermal pulse 
($T_{csh}^{MAX}$) depends on the mass of the H-exhausted core \citep[see, e.g.,][]{IR1983}:  
the larger the core mass, the larger $T_{csh}^{MAX}$.
Larger core masses are attained by more massive 
stars and, for a given mass, by more metal-poor stars. For instance,
at Z$=$0.0001, the \nean reaction provides an important contribution 
to the overall s-process nucleosynthesis for M$>$2.5 M$_\odot$, 
while at solar metallicity this limit rises up to 
3.5 M$_\odot$. 
At variance with the $^{13}$C burning, since the core mass
increases as the star climbs the AGB, the \nean reaction
is more efficient near the AGB tip, when the temperature within the
convective zone generated by the thermal pulse is larger.
  
Owing to the large temperature (T$> 300$ MK), the $^{22}$Ne burning generates 
quite large neutron densities, i.e., in between  10$^{11}$ and 10$^{13}$ 
neutrons/cm$^{3}$. As a consequence, some branchings, which are closed 
in the case of the radiative \ctan neutron burst, are opened, thus allowing 
alternative s-process paths. Nonetheless, the neutron exposure is 
not particularly high because of the  short duration 
of a thermal pulse. 
In addition, the iron reservoir in the convective zone generated by a thermal pulse
is large enough to guarantee 
a sufficiently large amount of seeds for the whole duration of $^{22}$Ne 
burning, even for metallicity as low as Z$\sim$0.0001. 
Then, the $f$ factor defined in the Introduction   
does not rich the large values attained in the case of the radiative $^{13}$C
burning as a consequence of the rapid Fe consumption. 
For all these reasons,
the  main yields of the second neutron burst are light-s elements, while 
heavy-s and Pb are marginally produced.
 
Note that at variance with massive stars, in AGB stars undergoing TDU 
the \nean reaction may be a primary neutron source. Indeed, the TDU 
moves primary C from the He-rich intershell to the H-rich envelope. 
In AGB models with M$< 4$ M$_\odot$ and low initial metallicity,
 we find that the C abundance in the envelope attains solar values 
just after a few TDU episodes. As a result, a relatively high amount of primary $^{22}$Ne 
is piled up within the He-rich layer. In any case, the amount of primary $^{22}$Ne in the intershell
decreases as the stellar mass increases.  
Indeed, as already discussed, HBB and HTDU prevent deep TDUs 
in the more massive AGB models, thus reducing the contamination of the envelope with the ashes
of the internal nucleosynthesis. For instance, we find that
the s-process contribution is substantially reduced in models
with M$> 6$ M$_\odot$. Such a conclusion confirms previous finding by \citet{doherty2014}. 
In their super-AGB models most of the neutrons released by the \nean reaction 
are captured by the $^{25}$Mg(n,$\gamma)^{26}$Mg reaction.
As a result, a marginal contribution 
to the synthesis of heavy elements is expected from super-AGB stars.

\section{The s-process polluters}\label{polluters}

In this section we present a set of models of AGB stars with mass 
$2\le M/M_\odot \le 6$, [Fe/H]=-1.7, [$\alpha$/Fe]=0.5 and Y=0.245\footnote{
The initial composition has been obtained following the procedure described
in \citet{piersanti2007}. The reference solar abundance compilation 
is from \citet{lodders2003}.}. The corresponding metallicity
is Z=7$\times 10^{-4}$.

Table \ref{tabphys} reports various properties of these models. In principle,
owing to the relatively short lifetime (second column of Table \ref{tabphys}), 
these stars could have time to evolve up to the AGB, thus contaminating the 
still not-completed building blocks of the early Galaxy.
In colum 2 and 3 we report the number of TDU episods and the total mass of the 
material dredged up, respectively. Note that the average mass dredged up in a single 
TDU episode decreases as the stellar mass increases. 
On the contrary, the number of TDU epsidodes increases
as the stellar mass increases. As a result the M$_{TDU}$ versus M relation is not linear.
In particular, M$_{TDU}$ is practically the same in all models with $3\le$M$\le5$,
while in the 6 M$_\odot$ model it is the 27\% of that of the 2 M$_\odot$ model.     
As it is well known, the maximum peak temperature in the convective
shells generated by thermal pulses (column 7) as well as 
the maximum temperature developed at the base of the
convective envelope during the interpulse periods (column 8) increase as the 
stellar mass (or the core mass) increases \citep{IR1983}. 
The first temperature affects the efficiency of 
the convective neutron-capture nucleosynthesis as powered by the \nean reaction, while the second 
temperature determines the efficiency of the HBB. 
We also report the maximum neutron density corresponding to the maximum peak temperature
attained at the base of the
convective shells generated by TPs (last column in Table 1). Note that, 
in spite of the similar temperature,
the maximum neutron densities of the 5.5 and 6 M$_\odot$ models 
are smaller than that of the 5 M$_\odot$. Such an
 occurrence is due to the limited dredge up suffered by the more massive AGB stars 
 and, in turn, to the smaller amount of primary $^{22}$Ne accumulated in the 
 He-rich intershell (see section 3.2). 

\begin{figure}
\includegraphics[angle=0,width=\columnwidth]{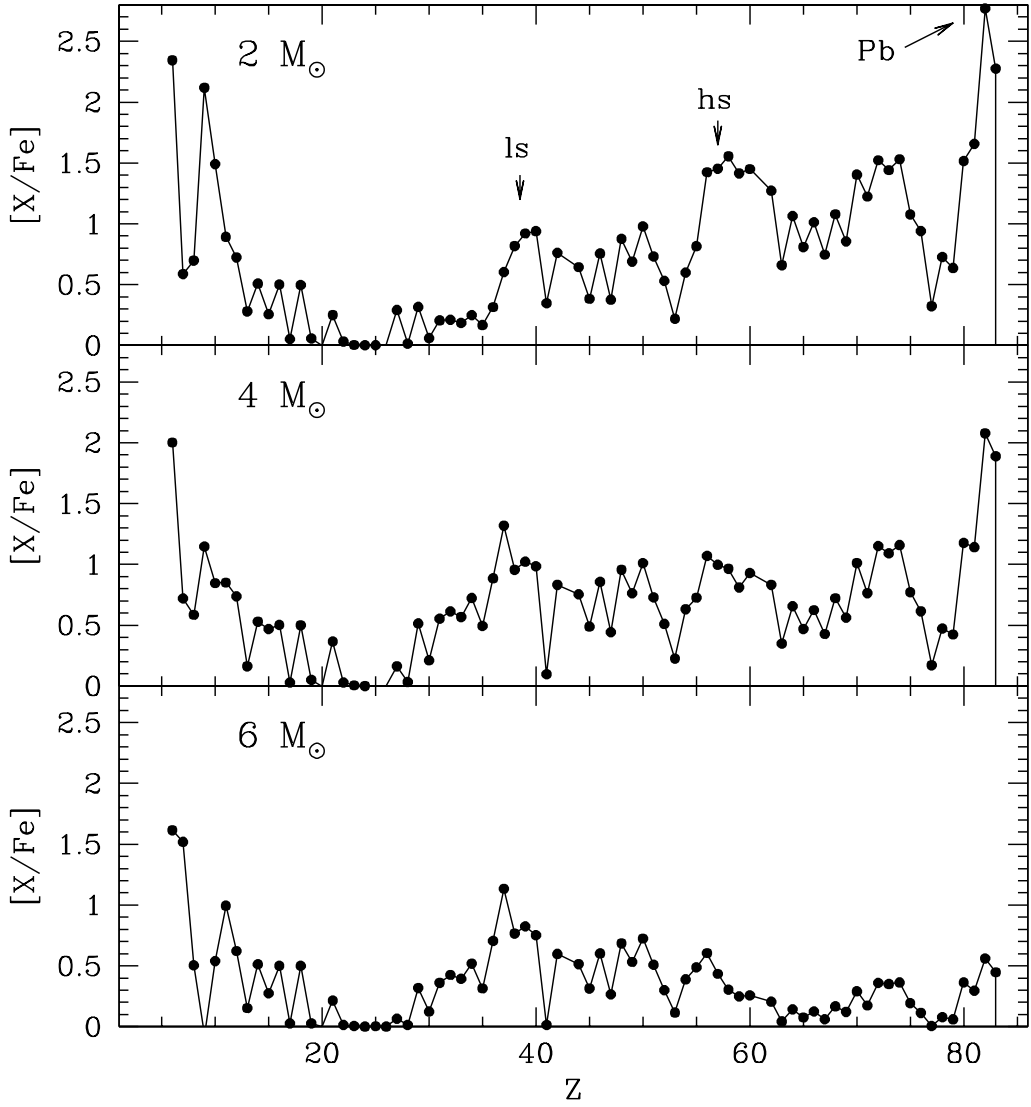}
\caption{Final compositions 
of the 2, 4 and 6 M$_\odot$ models with [Fe/H]=-1.7, [$\alpha$/Fe]=0.5 and 
Y=0.245.}\label{xfe}
\end{figure}

In  Figure \ref{xfe} we compare the final surface composition of three
of the computed models. In the upper panel we show the 2 M$_\odot$ case. This model
represents a typical case of a low-mass AGB stars, characterized by a low 
core mass, quite deep TDU episodes, low temperature at the base of the convective
zone generated by a TP and low temperature at the base of the
convective envelope. The nucleosynthesis is 
dominated by the radiative $^{13}$C burning, while the convective 
$^{22}$Ne burning only plays a marginal role and the HBB is negligible
\footnote{Measurements of C and N in Carbon  Enhanced 
Metal Poor  stars \citep[e.g.,][]{johnson2007}
demonstrate that some kind of extra-mixing, deep mixing or cool bottom process
should be active in these stars \citep[see also][]{cristallo2007}. 
This phenomenon, whose physical origin is a controversial issue, is not 
included in the models here presented. Note, however, that it does not
affect the synthesis of heavy elements.}. The opposite situation is 
illustrated in the lowest panel, where the 6 M$_\odot$ case is shown.
In this model the $^{13}$C pockets are rather small and the corresponding 
contribution to the s-process nucleosynthesis is marginal. On the contrary, 
due to the large core mass, the temperature at the base 
of the convective shell generated by a TP is quite large and the s process
powered by the \nean reaction is very efficient. These stars also experience a
substantial HBB, as clearly shown by the large nitrogen 
 enhancement. The central panel shows an intermediate case, the 
4 M$_\odot$ model. In these stars the synthesis of heavy elements 
is determined  by a combination of both the \ctan and the 
\nean neutron bursts. During the first part  of the TP AGB phase, 
the  $^{13}$C pockets are sufficiently large to power  a substantial  
production  of lead and, to a less extent, of heavy-s elements, while the lack of
$^{22}$Ne hampers the second neutron burst. 
After a few TPs,
physical and chemical conditions are reversed. The $^{13}$C pockets 
becomes progressively smaller, while the amount of $^{22}$Ne 
in the He intershell grows up as a consequence
of the C dredge up. 
Then, in the late part of the AGB, the convective $^{22}$Ne
burning dominates the s-process nucleosynthesis. The synthesis of lead, 
which is very efficient during the first part of the thermal pulse AGB evolution, reduces significantly 
after a few TPs, while the production of the light-s elements increases. 
Note the Rb peak (Z=37) in both the 4 and the 6 M$_\odot$ models. This feature is the signature 
of the high neutron density characterizing the $^{22}$Ne neutron burst, which is absent in models with M$\le 2.5$ M$_\odot$, 
because of the marginal activation of the second neutron source.  

In Table \ref{tabchim} we report the average overabundances in the ejected material
of some representative elements, namely:

\noindent
1) C+N+O. As a consequence of both the first and the second dredge up,
 material processed by the H burning is mixed into the envelope. As a result, the abundances of 
C and O decrease, while that of N increases. Nevertheless,
the total amount of C+N+O nuclei is conserved. This is not the case of the TDU, which
moves into the envelope primary C and, to a less extent, primary O produced by the He burning.
As a result, the overabundance of C+N+O at the end of the AGB depends on the
efficiency of the TDU and, then, it reflects the variation of 
M$_{TDU}$ with the initial mass (see Table \ref{tabphys}). 

\noindent
2) Fluorine. The AGB production of this element is strictly connected to the 
\ctan reaction \citep[][and references therein]{lugaro2004,abia2009}. Indeed, $^{19}$F is
mainly synthesized by the $^{15}$N$(\alpha,\gamma)^{19}$F reaction in the convective zone
generated by a TP. $^{15}$N may be produced by the 
$^{18}$O$(p,\alpha)^{15}$N
reaction, which requires the simultaneous presence of certain amounts of 
both protons and $^{18}$O. This condition is 
fulfilled in the $^{13}$C pocket, where protons are released by the
$^{14}$N$(n,p)^{14}$C reaction, the main neutron poison, and $^{18}$O mainly by the 
$^{14}$C$(\alpha,\gamma)^{18}$O reaction.
If the C+N+O abundance in the envelope is large enough, an additional source of
$^{15}$N is provided by the $^{13}$C left in the H-burning ashes and  
engulfed in the convective zone generated by a TP. At the beginning of the
thermal pulse, this $^{13}$C is rapidly consumed by the \ctan reaction, thus allowing the
production of  $^{15}$N through the same nuclear chain already 
active in the $^{13}$C pocket, with the additional contribution of the 
$^{14}$N$(\alpha,\gamma)^{18}$F($\beta)^{18}$O reaction.
In a low-mass AGB star with solar Z, 
more than 50\% of the fluorine enhancement is due to the $^{13}$C left in the H-burning ashes, 
while at low Z this contribution becomes important in the late part of the AGB, 
when, as a consequence of the TDU, the C+N+O in the envelope increases. By increasing the 
stellar mass, the F production decreases, according to the progressive reduction of the
$^{13}$C pockets mass and of the TDU. In addition, owing to the larger temperature within 
the convective zone generated by a TP, the two reactions $^{19}$F$(\alpha,p)^{22}$Ne and 
$^{19}$F$(n,\gamma)^{20}$F (followed by a 
$\beta$ decay into $^{20}$Ne) become efficient fluorine destroyers. Finally, 
in the more massive AGB a further depletion of F is due to the HBB.  
Since also Pb is a main product of low-mass stars,
while it is underproduced in the more massive AGB, a positive correlation between 
F and Pb is expected in s-rich GC stars.

\begin{figure}
\includegraphics[scale=1.0,angle=0] {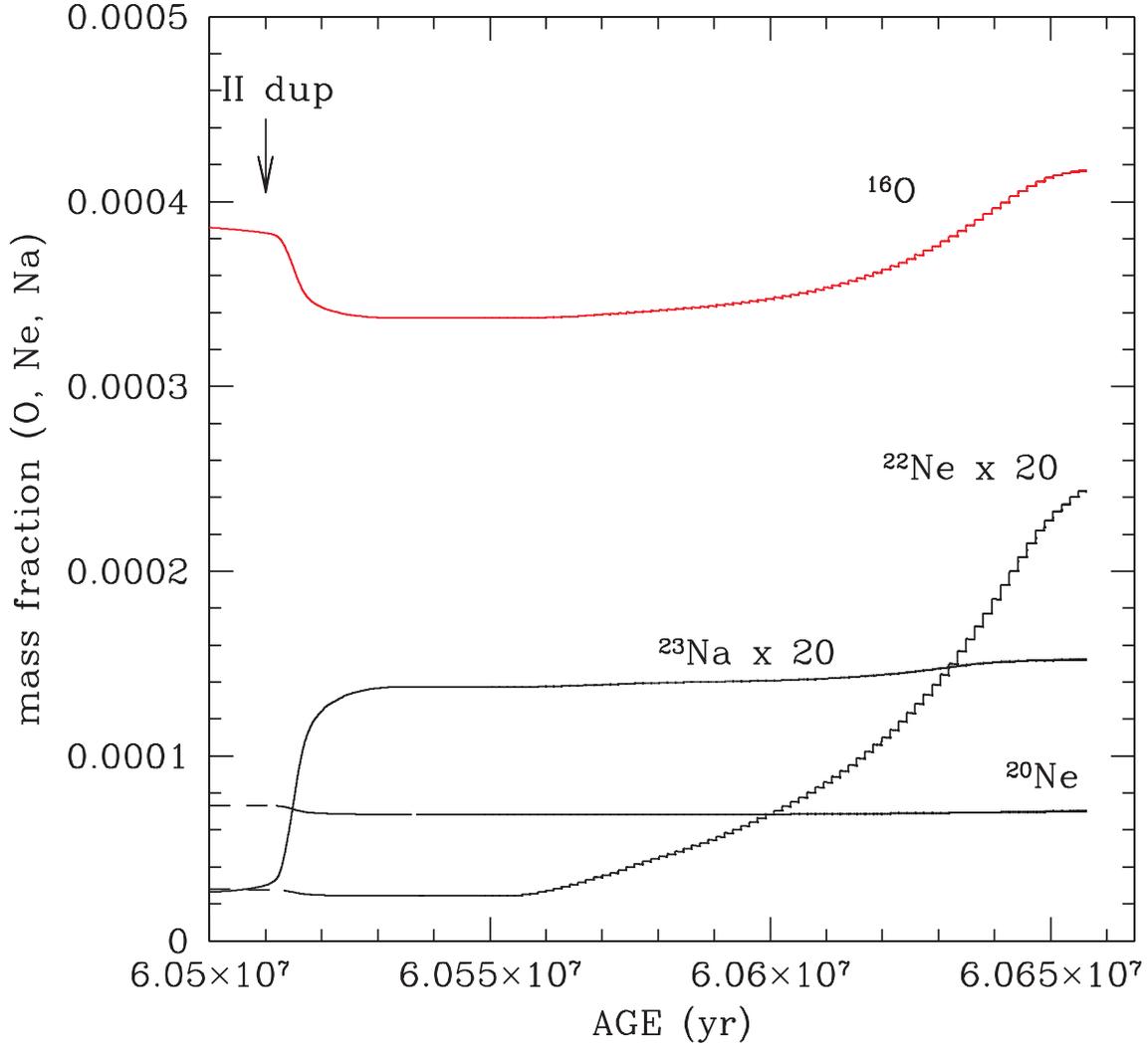}
\caption{Evolution of $^{16}$O, $^{20}$Ne, $^{22}$Ne and $^{23}$Na at the surface of the 
6 M$_\odot$ model. $^{16}$O is depleted after the second dredge up and, later on, it is
restored to nearly the initial abundance by the TDU, which attain the region 
enriched with primary O during the preceding TP. The two major Ne isotopes are not affected 
by the second dredge up. However, $^{22}$Ne abundance increases after each TDU.
$^{23}$Na is enhanced after the second dredge up, while a negligible increase 
is observed as a consequence of the TDU. Finally, no one of these isotopes 
are affected by the HBB.}\label{Na}
\end{figure}

\noindent
3) Sodium. Na partecipates to the Ne-Na cycle active in the hottest zone of the 
H-burning shell. In low-mass AGB stars a further sodium source is provided by the so-called 
$^{23}$Na pocket \citep{goriely2000, cristallo2009}, as due to an incomplete Ne-Na cycle occurring
in the thin layer with variable H profile left by the third dredge up.
In addition, neutron captures on $^{22}$Ne may produce some
$^{23}$Na within the He intershell. Therefore, the surface sodium abundance increases 
after each dredge-up episode. 
In the more massive AGB, the HBB modifies 
the surface sodium abundance. It may be produced or destroyed, depending on the
maximum attained temperature and on the interplay with the TDU that brings in the envelope 
additional $^{22}$Ne \citep[see][]{ventura2006}.
The large part of the Na enhancement we find in our models with M$>3$ M$_\odot$ is
a consequence of the second dredge up, while the HBB and the TDU play a marginal role
(see Figure \ref{Na}).
In lower mass models, additional contributions come from the third dredge up.
Owing to the combination of different sources, no clear correlations between the Na 
abundance and the stellar mass can be derived.

\noindent
4) Y, La and Pb. These three 
elements are representative of light-s, heavy-s and of the end point of the s-process
nucleosynthesis, respectively. Relative abundances are also reported in the last two 
columns. As expected, lead is the major product of low-mass AGB. Its overabundance decreases as 
the mass increases. In the more massive models it is underproduced with respect to 
the light-s. On the contrary, light-s elements are underproduced by 
low-mass stars, while for M$\ge 3.5$ M$_\odot$ they are as overabundant as 
the heavy-s or even more abundant. As for the Pb, 
the La overabundance decreases as the mass increases. 

\begin{figure}
\includegraphics[scale=1.0,angle=0]{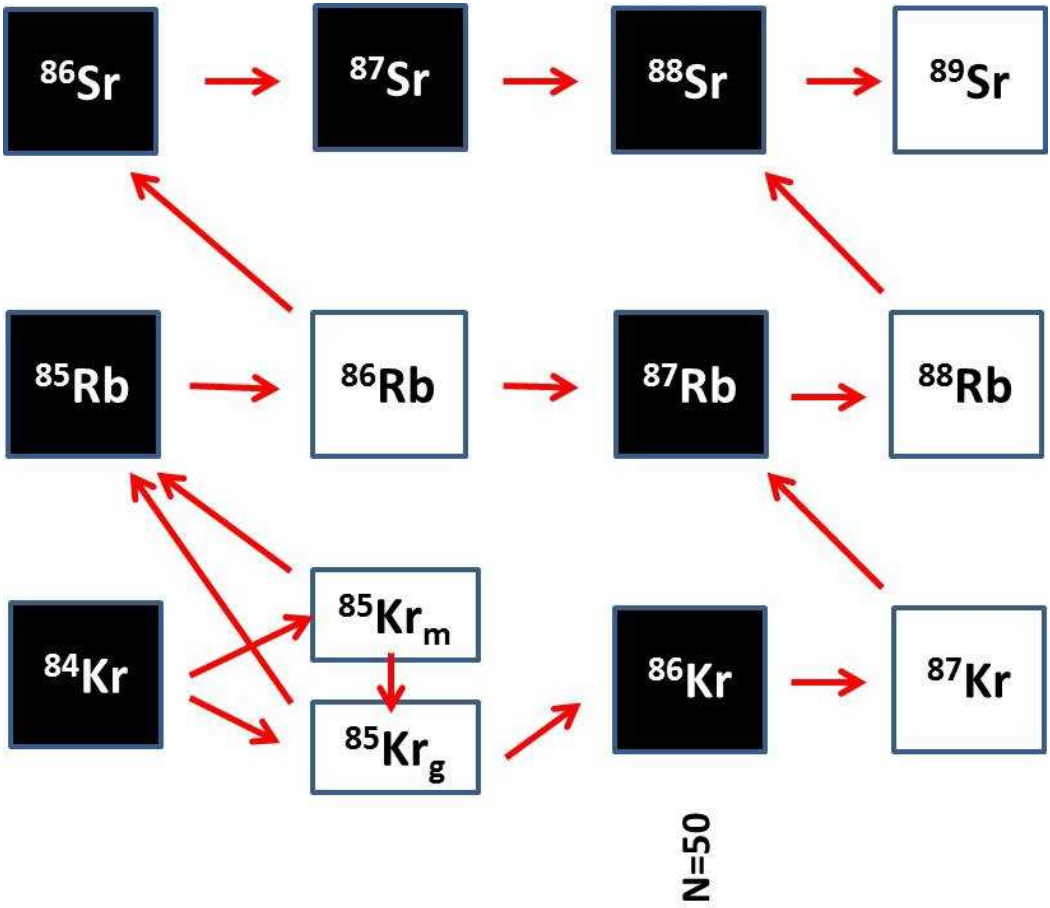}
\caption{The s-process path in the Kr-to-Sr region. Empty squares represent
unstable nuclei, while arrows show the alternative s-process paths}\label{kr85} 
\end{figure}

\noindent
5) Rubidium. As already recalled, the overabundance
of this element with respect to the other light-s, such as Sr, depends on 
the neutron density developed in the convective 
\nean neutron burst. The nuclei chart in the Rb region is shown in Figure 
\ref{kr85}. In the radiative \ctan neutron-capture nucleosynthesis, 
the two major branchings at $^{85}$Kr and $^{86}$Rb are practically closed, so that
the s-process path proceeds through the sequence 
$^{84}$Kr-$^{85}$Kr-$^{85}$Rb-$^{86}$Rb-$^{86}$Sr and, then, to the neutron magic $^{88}$Sr.
As a result, the [Rb/Sr]$<0$. The case of the  \nean neutron burst is substantially different.  
About 50\% of the neutron captures on $^{84}$Kr directly produce ground state $^{85}$Kr,
whose $\beta^-$ decay half-life is 10.756 yr. 
The remaining 50\% feeds the isomeric state of $^{85}$Kr (half-life 4.480 h),
which decays $\beta^-$ (78.6\%) and $\gamma$ (21.4\%).
As a result, about 60\% of the neutrons captured by $^{84}$Kr produce $^{85}$Kr$_g$.
Since the temperature at the base of the
convective zone generated by a thermal pulse remains above 300 MK for no more than 1 yr
(the precise value depends on the mass), $^{85}$Kr$_g$ practically behaves as a stable 
nucleus during the whole neutron-capture nucleosynthesys episode. 
Nonetheless, it may capture a 
neutron producing the neutron magic $^{86}$Kr. Then, a further neutron capture
would finally lead to $^{87}$Rb. However, the latter is hampered by the low 
MACS of $^{86}$Kr, namely $3.4\pm 0.3$ mbarn\footnote{
The values of the cross sections here reported are the 30 KeV MACS from the KADONIS database, 
\citep{kadonis2009}.}. Indeed, we find that 
as a consequence of this s-process path, $^{86}$Kr is accumulated rather than $^{87}$Rb
\citep[see also][]{vanraai2012}. Nevertheless, the $^{85}$Kr$_g$ remained unburnt at the end 
of the TP will decay into $^{85}$Rb, and, later on, it will be 
dredge up to the envelope. Note that the efficiency of this Rb source depends on the rather 
uncertain neutron capture cross section on $^{85}$Kr$_g$ (see section \ref{discussion}).        
The 40\% of the $^{84}$Kr$(n,\gamma)$ reactions produces  $^{85}$Kr$_m$ that 
suddenly decays $\beta$ into $^{85}$Rb and, after a further neutron capture, 
leads to $^{86}$Rb. Due to the large cross section ($202\pm 163$), the neutron capture on 
$^{86}$Rb competes with the $\beta$ decay (half-life 18.63 d), so that a certain
amount of the neutron magic $^{87}$Rb can be produced. Also in this case, 
the precise evaluation of the efficiency of this Rb source relies on the rather uncertain 
MACS of 
the $^{86}$Rb$(n,\gamma)$ reaction. We have performed some tests to distinguish the various 
contributions to the Rb synthesis. These tests confirm previous finding of 
\citet{vanraai2012}, namely the dominant contribution is from the $^{86}$Rb branching.
   
\begin{figure}
\includegraphics[angle=0,width=\columnwidth] {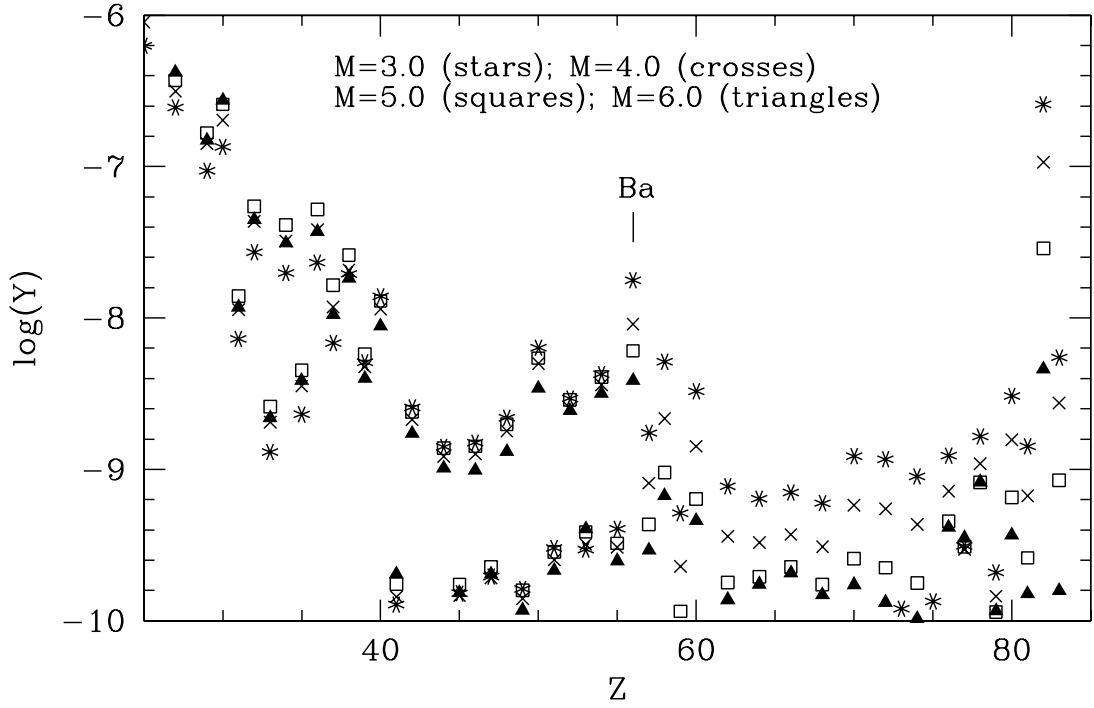}
\caption{Total yields, i.e., the total mass of a given element (in M$_\odot$) 
in the material ejected by the star during its whole life.  
Results for models with initial mass 3, 4, 5 and 6 M$_\odot$ are here compared.
}\label{yields}
\end{figure}

The heavy element yields (from Fe to Bi) of the 3, 4, 5 and 6 M$_\odot$ models are compared 
in Figure \ref{yields}. Below the Ba peak the yields of the various models are very similar 
(within a factor of 2), except for the 3 M$_\odot$ that shows a smaller overproduction
of  all the elements usually ascribed to the weak s-process component($Z<36$).
Above the Ba peak, the differences are definitely more pronounced. 
If the total La mass ejected by the
3 M$_\odot$ is about 5 time larger than that ejected by the 6 M$_\odot$, the Pb ejected mass is
about 30 time larger. As already stated, the more massive AGB stars 
provide a little contribution to the synthesis of the heaviest s elements. 

\begin{figure}
\includegraphics[angle=0,width=\columnwidth] {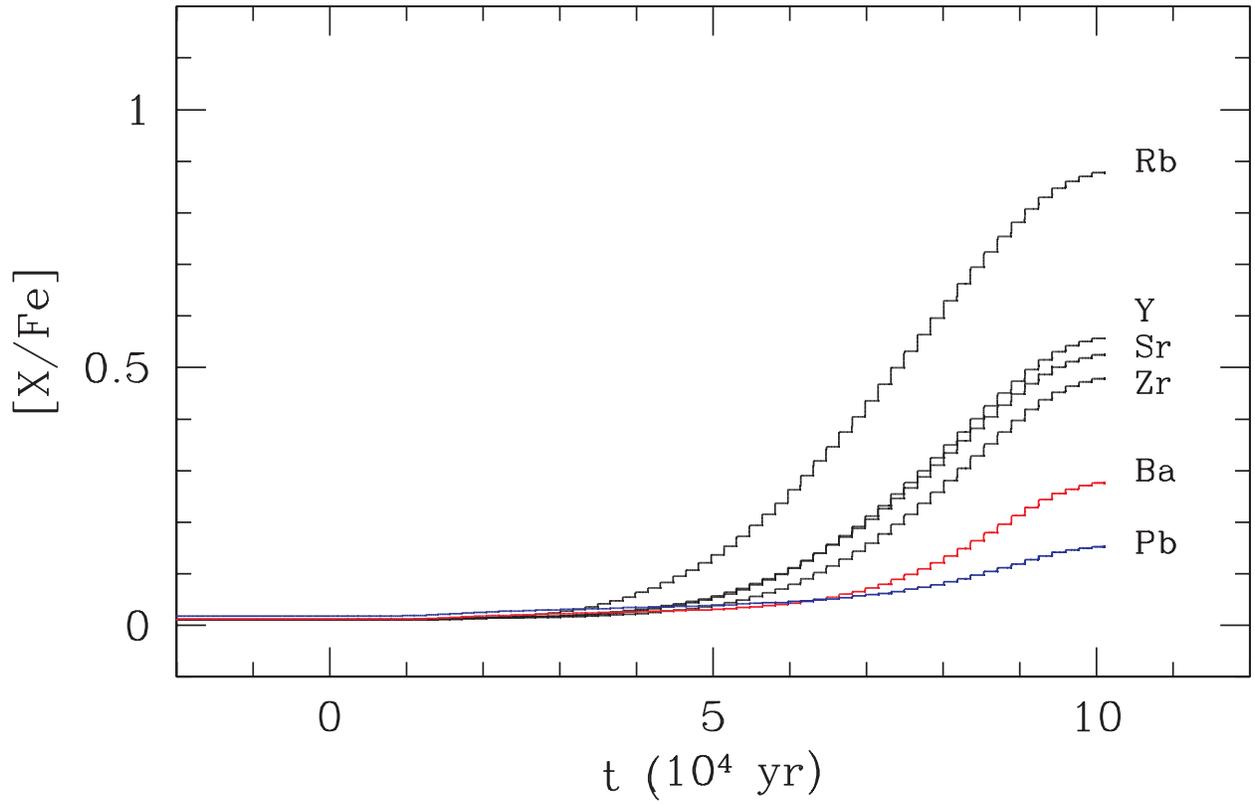}
\caption{Evolution of the s process surface abundances for a model of M=6 M$_\odot$ and Z=0.001.
}\label{m6z001}
\end{figure}

As far as we know, our stellar models are the only ones obtained by 
coupling the stellar structure equations with a full network of 
chemical evolution equations. Indeed, s-process calculations in AGB stars 
are usually obtained by means of post-process codes 
\citep{gallino1998, goriely2000, lugaro2012}.
Probably the main difference concerns the inclusion of the $^{13}$C pocket that in 
post-process calculations is decoupled from the evolution of the physical structure.
Indeed, in the extant post-process calculations the same $^{13}$C pocket is assumed for 
the whole evolutionary sequence. As shown by \citet{cristallo2009}, this is a rather 
crude approximation, because of the natural shrinking of the He-rich intershell 
and, in turn, of the $^{13}$C pocket, due to
the increase of the pressure gradient caused by the increase of the core mass 
occurring during the AGB evolution.       
Comparisons between our nucleosynthesis results for low-mass stars 
and those obtained by other authors can be found in several papers. 
\citet{cristallo2009} find a qualitative agreement with the s-process calculations
of \citet{gallino1998}. Note that the latter were based on
old stellar models \citep{straniero1997}, obtained by using an essential 
nuclear network, Reimers mass loss and neglecting the modifications of 
the radiative opacity caused by the TDU. 
More recently, \citet{lugaro2012}
present new post-process calculations based on updated stellar models with Z=0.0001.
In spite on the many differences in both stellar models and nucleosynthesis, 
their results for low-mass stars are in very good agreement 
with those we have reported in \citet{cristallo2009}, for both light and heavy elements 
(see table 5 and 6 in \citet{lugaro2012}). Concerning more massive AGB stars, 
the  models here presented have an initial composition quite different from that 
of the \citet{lugaro2012} models. 
In particular, the iron abundance (the s-process seed) is about 3 times larger in our models.
Recently, \citet{dorazi2013d} presented two more models 
of massive AGB stars, 5 and 6 M$_\odot$, obtained with the same stellar 
evolution code of \citet{lugaro2012} \citep[see also][]{karakas2010},
with total metallicity Z=0.002 and [$\alpha$/Fe]$=0.4$. 
This composition is more similar to that of 
a model of 6 M$_\odot$ we have recently computed with our FUNS code 
for a work in progress on galactic chemical evolution. 
The initial iron content is the same as in D'Orazi et al., even if we assume 
[$\alpha$/Fe]$=0$, so that Z$=0.001$.
Note that D'Orazi et al. present two models computed under different assumptions 
for the mass-loss rate, namely \citet{vw1993} and  \citet{blocker1995}, respectively.
Like us, they use radiative opacity tables that account for the
effects of the TDU (see section 2.5).   
The evolution of the surface abundances of representative 
s-process elements of our model are shown in Figure \ref{m6z001}. 
This plot can be directly compared with Figure 11 in the D'Orazi et al. paper. 
In spite of the many differences of the two stellar models, 
the overall result appear quite similar. 
In particular, both calculations show a significant enhancement of  Rb 
and, to a less extent, of other light-s elements (Sr, Y and Zr), 
while heavier elements are marginally produced. 
Note that D'Orazi et al. neglect the s-process contribution due to the radiative 
\ctan burning. The low Pb abundance we find confirms that this is a good 
approximation for the more massive AGB stars. 
Nevertheless, they use the \nean rate reported in the \citet{NACRE} compilation, 
while we use the more recent \citet{jager2001} that is about 50\% smaller
at the temperatures of the shell-He burning (about 350 MK). 
Some minor discrepancies in the resulting surface composition probably reflect such a difference.
The TP-AGB lifetime we find is intermediate between those obtained by
D'Orazi et al. with the two different choices of the mass-loss rate. 
Indeed, the mass-loss rate we use is intermediate between \citet{vw1993}
and \citet{blocker1995} (see section 2.1). Consequently,
our final heavy element abundances are in between the two obtained by D'Orazi et al.   
Once the initial composition is properly re-scaled (our model is for [O/Fe]=0), 
the final C+N+O of our models is also consistent with the differences in the mass-loss rate. 
It implies a similar TDU efficiency. 
On the contrary, the HBB is less efficient in our model. 
Indeed, at variance with D'Orazi et al., in our model, the C dredged up is only 
partially converted into N, while O, Na, Mg and Al are very marginally affected by the HBB. 
This discrepancy may be partially attributed to different input physics, such as the EOS, 
interpolation on radiative opacity tables, super-adiabatic convection and the like. 
However, the use of different mixing schemes may be the main origin of this difference.   
In particular, at variance with D'Orazi et al., who  assume 
instantaneous mixing, we make use of a time-dependent mixing scheme 
(see section 2.3).                                         

\section{Multiple populations and heavy elements in Globular Clusters}\label{multipop}
A growing amount of observational evidences, among which multiple 
sequences in color-magnitude diagrams, extremely blue horizontal branches, 
cyanogen variations and their anticorrelation with CH, 
O-Na and Mg-Al anticorrelations, support multiple population (MP) scenarios 
for the formation of GCs 
\citep[][and references therein]{cottrell1981,dantona2005,gratton2012}.
Alternative models, such as the accretion on main-sequence 
low-mass stars of material lost by more massive objects, have been also 
proposed \citep{dantona1983,dantona2002}.
Our aim is to verify if these models can also account for the observed
s-process pollution and under which conditions this chemical
anomaly arises.
In the rest of the paper we discuss the case of 
MP models, even if the yields presented in the 
previous section may be also used to test accretion models.  

\begin{figure}
\includegraphics[angle=0,width=\columnwidth]{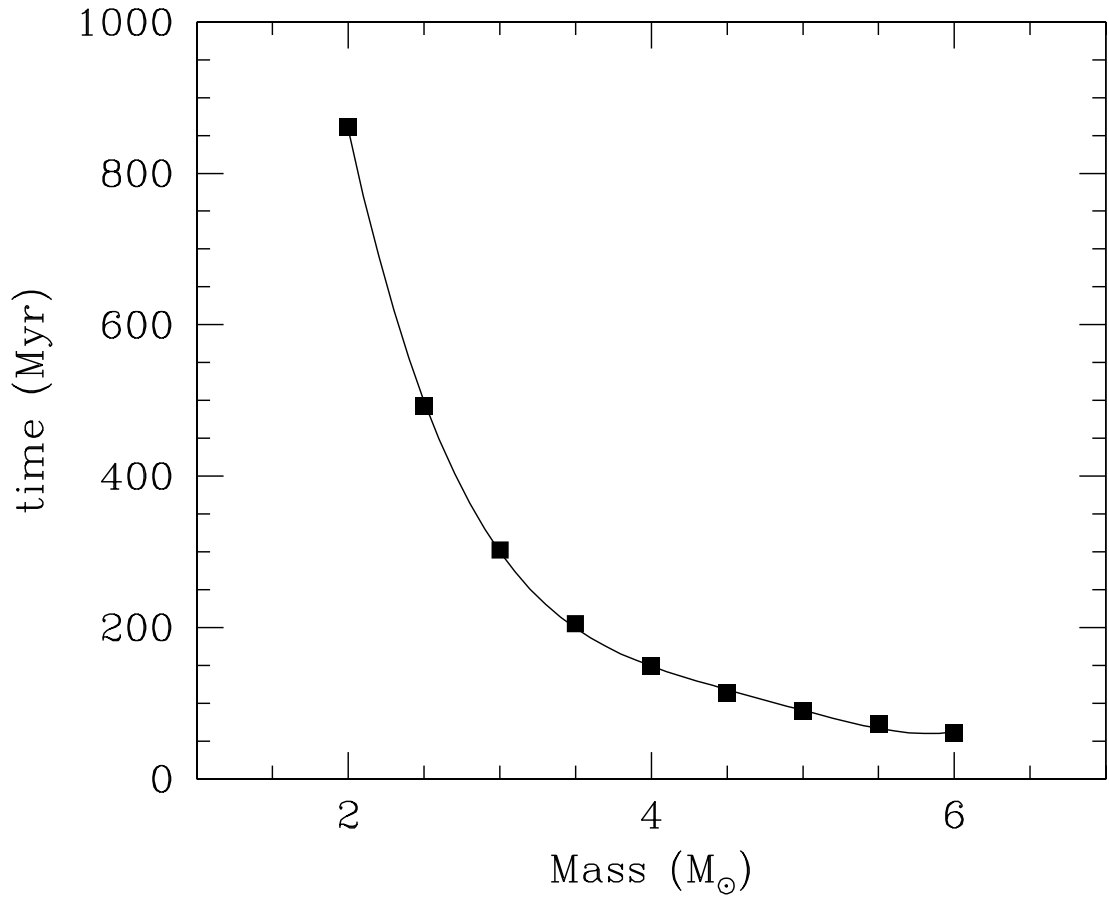}
\caption{Stellar lifetime versus initial mass.}\label{mtime}
\end{figure}

MPs may be the result of i) multiple star formation episodes within the same
cluster or ii) merging of smaller stellar systems containing a single stellar 
population. In principle, a combination of the two types is also possible. In any case,
the pollution responsible for the observed chemical variations may be
due to external stellar populations (primordial-pollution) 
or caused by the same stars of the cluster (self-pollution). 
In the following, we will assume that the polluters are 
normal halo stars, i.e., they form from s-poor gas with low [Fe/H],
[$\alpha$/Fe]$>0$ and [r/Fe]$>0$\footnote{ 
Note that [r/Fe]$=0$ in our stellar models. This assumption does not 
affect the physical evolution of a star. Nonetheless, when comparing
our nucleosynthesis predictions to the heavy elements
composition of GC stars, the r-process contribution should be properly subtracted
from the observed abundances.}. Moreover, we will not distinguish between 
primordial-pollution or self-pollution scenarios.
Hereinafter the term {\it first generation} refers to the polluters. 
As these stars evolve, they lose material containing the imprint 
of the internal nucleosynthesis. Possibly, this gas  
is mixed with some amount of residual pristine gas and, then, diluted. 
The total yield of the stars belonging to this first generation
with mass between $m$ and $m+\Delta m$ will be:
\begin{equation}
Y_j=\int_{m}^{m+\Delta m}\varphi (m)y_j(m)dm
\end{equation}
where  $\varphi (m)$ is the mass distribution function (MF) and $y_j(m)$ is
the mass of a given chemical species $j$ ejected by a star with mass $m$.
In the case of a power-law mass function, $\varphi (m)=A\times m^{-\alpha}$.
Then, keeping constant the value $y_j(m)=y_j(M_i)$ in the interval $M_i-0.25< m< M_i+0.25$\footnote{
Our models are spaced by 0.5 M$_\odot$.}, the total yield after a time 
$\Delta t$ is given by:
\begin{equation}\label{yield}
Y_j=\sum_{M_{min}(\Delta t)}^{M_{max}}y_j(M_i)\int_{M_i-0.25}^{M_i+0.25}\varphi (m)dm.
\end{equation}
$M_{min}$ and $M_{max}$ represent the minimum and the maximum 
initial mass of the stars that are expected to contribute 
to the s-process contamination of the interstellar gas. 
$M_{min}$ depends on the duration of the pollution phase
($\Delta t$): the larger $\Delta t$ the smaller $M_{min}$.
The initial mass - stellar lifetime relation, which is equivalent to 
the $M_{min}$-$\Delta t$ relation, is shown in Figure \ref{mtime}, where  
the solid curve represents a polynomial best fit:
\begin{equation}
\Delta t(Myr)=10.508\times M_{min}^4 - 199.62\times M_{min}^3 + 1422.5\times M_{min}^2 - 4561.5\times M_{min} + 5720.5
\end{equation}
In the following, we will assume $M_{max} = 6$ M$_\odot$, i.e., 
more massive stars do not contribute to the synthesis of 
s-process elements. As recalled in the Introduction, fast-rotating
massive stars might contribute to the weak component, but here we will limit 
our analysis to the main and the strong components. Morover, this assumption also implies that
we neglect possible contributions to the s process from super-AGB stars 
(see section \ref{22NE}). 
Then, if the total mass returned to the interstellar medium (m$_{agb}$) is
mixed to a certain amount of residual pristine gas (m$_p$), the resulting
mass fraction ($X_j$) is:
\begin{equation}
X_j=X_j^{agb}\frac{m_{agb}}{m_{tot}}+X_j^{p}\frac{m_{P}}{m_{tot}}=X_j^{agb} d+X_j^{p} (1-d)
\end{equation} 
where $X_j^{agb}$ and  $X_j^{p}$ are the mass fractions in the AGB ejecta\footnote
{$X_j^{agb}=Y_j/m_{agb}$, where $Y_j$ is the yield of equation \ref{yield}.} 
and in the pristine gas, respectively, 
$m_{tot}=m_{agb}+m_{p}$ is the resulting total mass and 
$d=m_{agb}/m_{tot}$ is the {\it dilution factor}.
Therefore, the overabundance with respect to iron will be:
\begin{equation}
\bigg[\frac{X_j}{Fe}\bigg]=\log\bigg(\frac{X_j^{agb}}{Fe} d + \frac{X_j^{p}}{Fe} (1-d)\bigg)-\log\bigg(\frac{X_j}{Fe}\bigg)_\odot.
\end{equation} 
Since we assume that the prestine gas is s-process free, 
$X_j^p$ is zero, so that:
\begin{equation}
\bigg[\frac{X_j}{Fe}\bigg]=\log\bigg(\frac{X_j^{agb}}{Fe} d\bigg)-\log\bigg(\frac{X_j}{Fe}\bigg)_\odot.
\end{equation}. 

Summarizing, the unknown quantities in this simple model are: $\Delta t$, the MF
of the first stellar generation and the dilution factor ($d$).
Note that all the constants, such as the total mass of the primordial stellar generation,
are canceled when relative abundances are considered. 

\begin{figure}
\includegraphics[angle=0,width=\columnwidth]{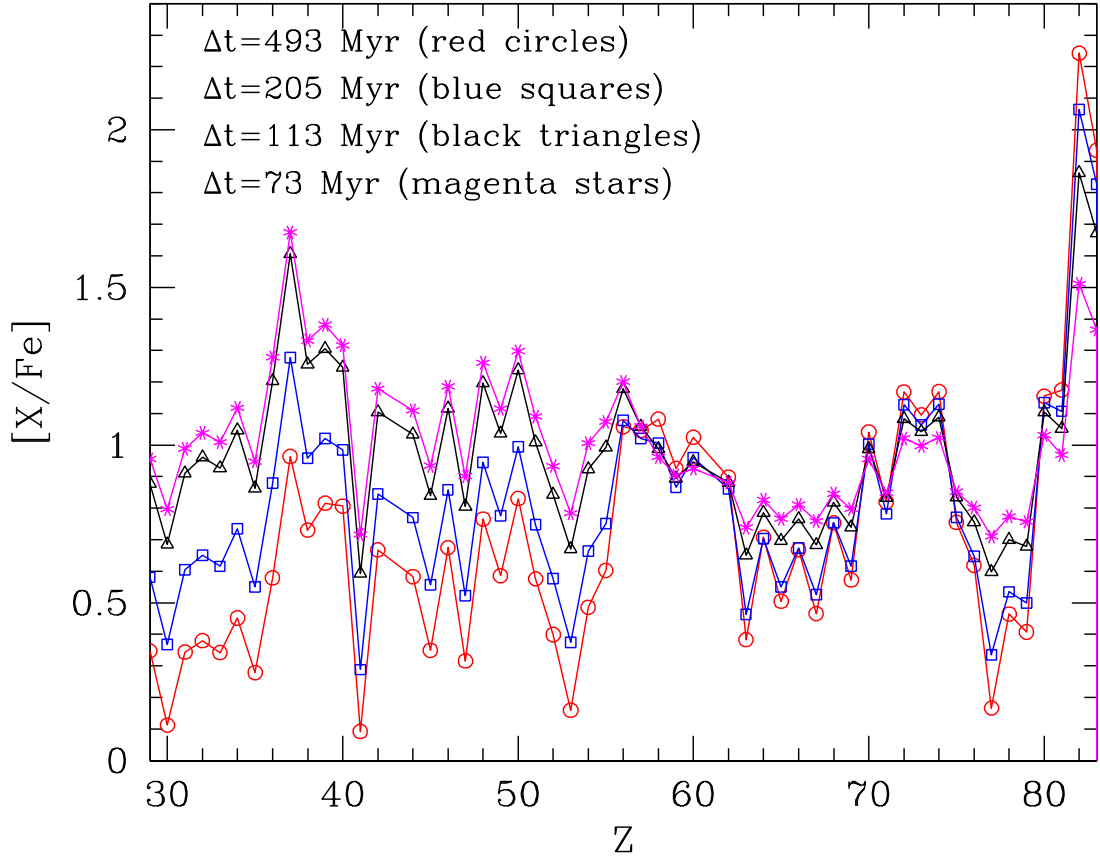}
\caption{Average composition of the material ejected by the first generation
AGB stars. The various curves refer to different delay times (see text). They
are shifted to the same [La/Fe].}
\label{xfeage}
\end{figure}

The resulting undiluted ($d=1$) compositions for different 
$\Delta t$ are compared in Figure \ref{xfeage}. Here, 
we have assumed a power-law MF with $\alpha=2.35$, namely 
a classical Salpeter mass function. As noted by \citet{kroupa2001}, this MF is a reasonable 
choice for M$\ge3$ M$_\odot$. 
The various curves are shifted in order 
to have the same [La/Fe] of the $\Delta t=493$ Myr case. 
As expected, light-s elements ($Z<40$) increase by decreasing $\Delta t$. In fact,
a short timescale implies more massive polluters that are dominated by the 
neutron-capture nucleosynthesis powered by the convective \nean reaction.
On the contrary, the abundances of Pb and Bi, mainly produced by the 
radiative \ctan nucleosynthesis, decrease as $\Delta t$ decreases.
In this way, the relative abundances between Pb, heavy-s and light-s elements
are potential indicators of the duration of the pollution phase.

\begin{figure}
\includegraphics[angle=0,width=\columnwidth]{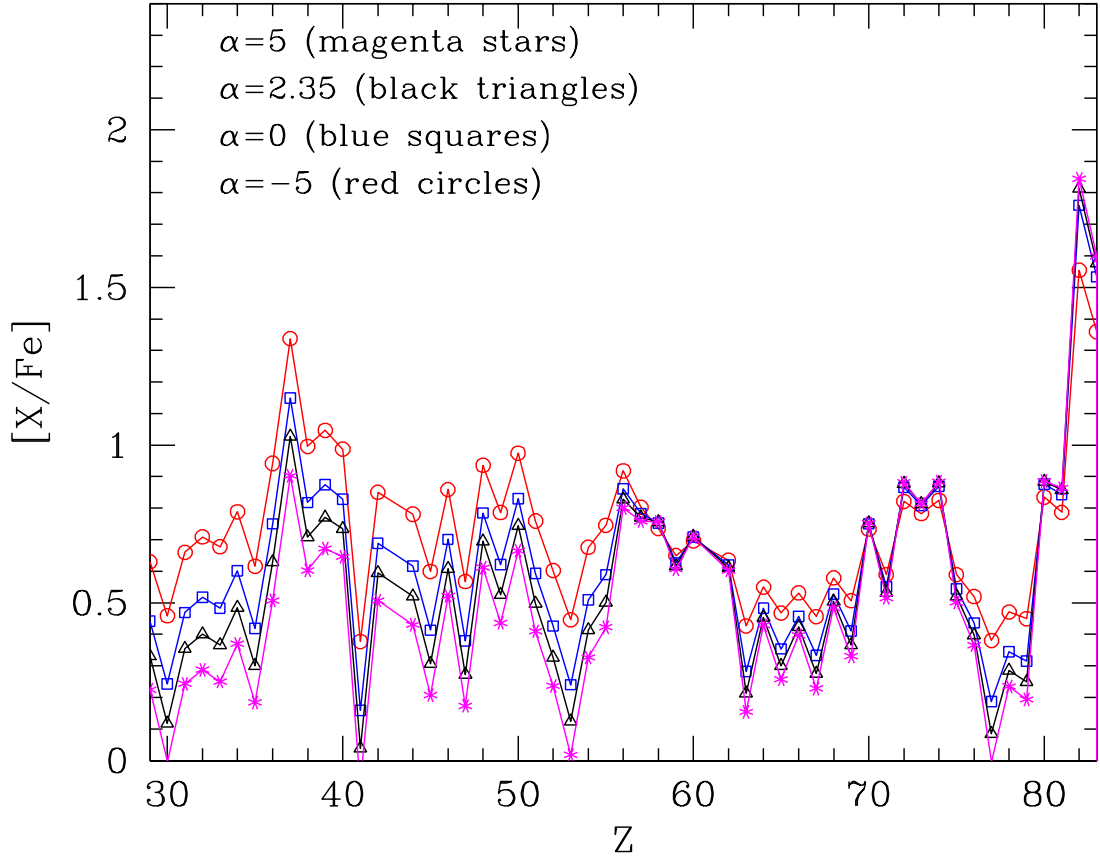}
\caption{Average composition of the material ejected by the first generation AGB
stars ($\Delta t=205$ Myr). The various curves have been obtained by varying the 
exponent of the power-law mass function. As in Figure \ref{xfeage}, they are
shifted to the same [La/Fe].}\label{xfemf}
\end{figure}

Finally, the effects of a variation of the MF are illustrated in Figure \ref{xfemf}.
All the models shown in this Figure have $\Delta t=205$ Myr, but different   
exponent of the power-law MF, namely: $-5\le \alpha \le 5$.
Note that positive $\alpha$ values correspond to the most common case of a MF that 
decreases as the stellar mass increases. On the contrary, negative  $\alpha$ values 
imply a MF that increases as the mass increases; in that case a maximum of 
the MF could be located above M$_{max}$. A similar MF has been
obtained by \citet{yoshii1986} \citep[see also][]{nakamura2002} in case of extremely 
metal poor environments. We are aware that this peculiar MF is in contrast with
many observational evidences as well as theoretical estimations of the GC mass function.
The models here reported should be considered as a test to check the sensitivity 
of our results on the adopted MF.
The various curves have been shifted to match the [La/Fe] value of the 
$\alpha=2.35$ case. Note that small $\alpha$ values favor the production 
of light-s elements, while the Pb abundance is suppressed.
In synthesis, if $\alpha\ge 0$ the differences implied by a change
of the MF are rather small. The (extreme) case $\alpha=-5$ mimics a moderate 
reduction of the $\Delta t$.

\section{Discussion}\label{discussion}
In the previous sections we have derived the chemical contamination
of pristine gas caused by a first generation of intermediate mass polluters. 
In particular we have shown how the 
composition of the gas returned to the interstellar medium
is expected to change as a function of the 
duration of the pollution phase that precedes the formation of the stars
presently showing an altered heavy-element composition.

In order to test such a theoretical prediction,  we have compared
the predicted chemical pattern to the observed s-process composition of M22 
and M4 stars. For these two clusters, extended samples 
of heavy element abundances are available. 
All the stars of M4 so far analyzed present similar overabundances of 
the s-elements  \citep{yong2008}, while only a subset of M22 members    
are s rich \citep{roederer2011}. This difference is likely the result of a different
formation history. Nevertheless, as illustrated in the previous section, 
our model may be applied to both primordial and self-pollution scenarios. 
Note that the stellar models here used have been computed for 
[Fe/H]=-1.7. This is quite similar to the value estimated for the first stellar 
generation of M22, i.e., [Fe/H]$=-1.82\pm 0.02$ \citep{marino2012}, but smaller than 
[Fe/H]$\sim-1.2$ attributed to M4 \citep{yong2008}.
The original iron abundance mainly affects the contribution to the s process due to the
radiative \ctan burning. In particular, a reduction of the Pb production is 
expected by increasing [Fe/H].  

\begin{figure}
\includegraphics[angle=0,width=\columnwidth]{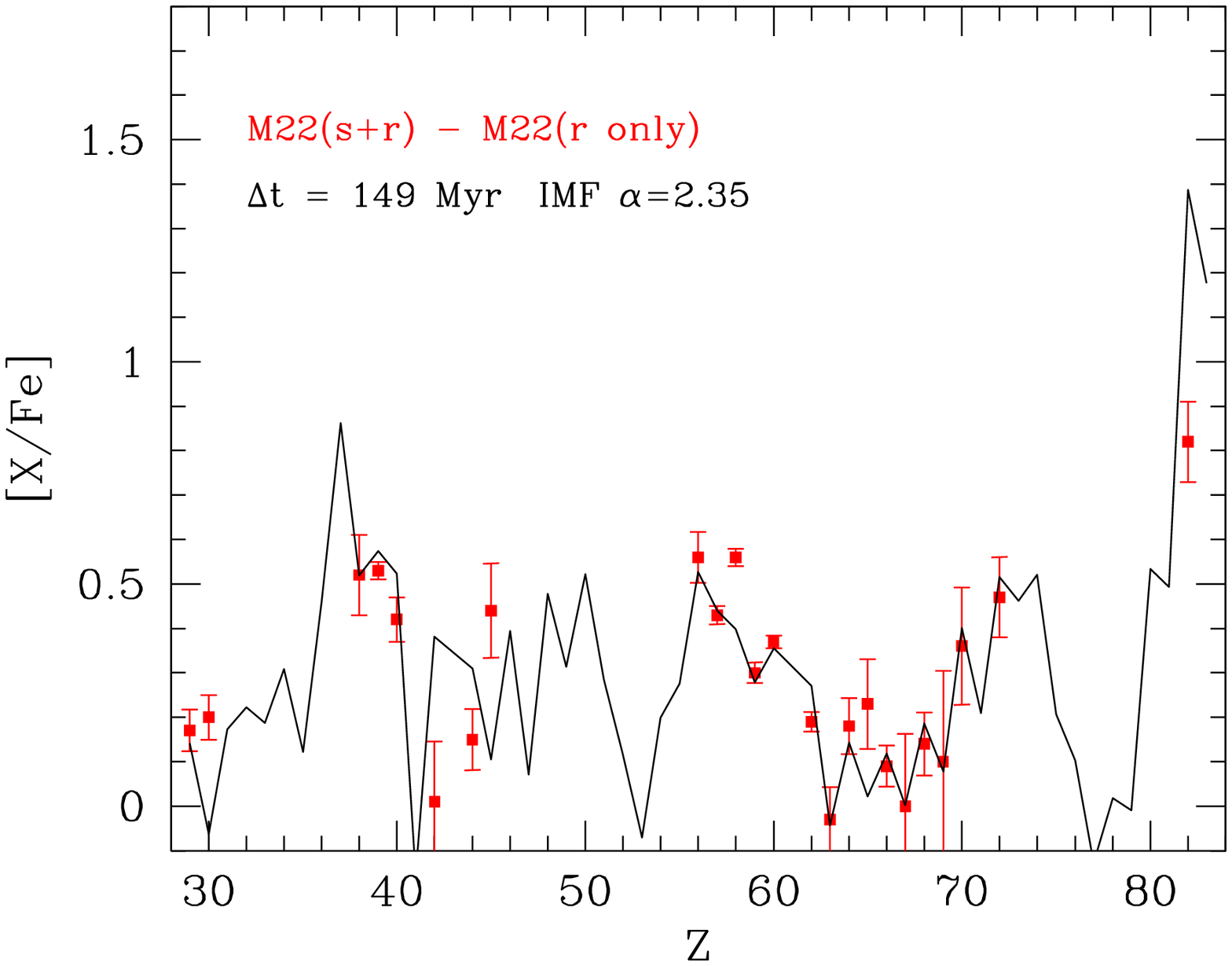}
\caption{Best fit of the average s-process chemical pattern of 
stars in M22.}\label{M22}
\end{figure}

\begin{figure}
\includegraphics[angle=0,width=\columnwidth]{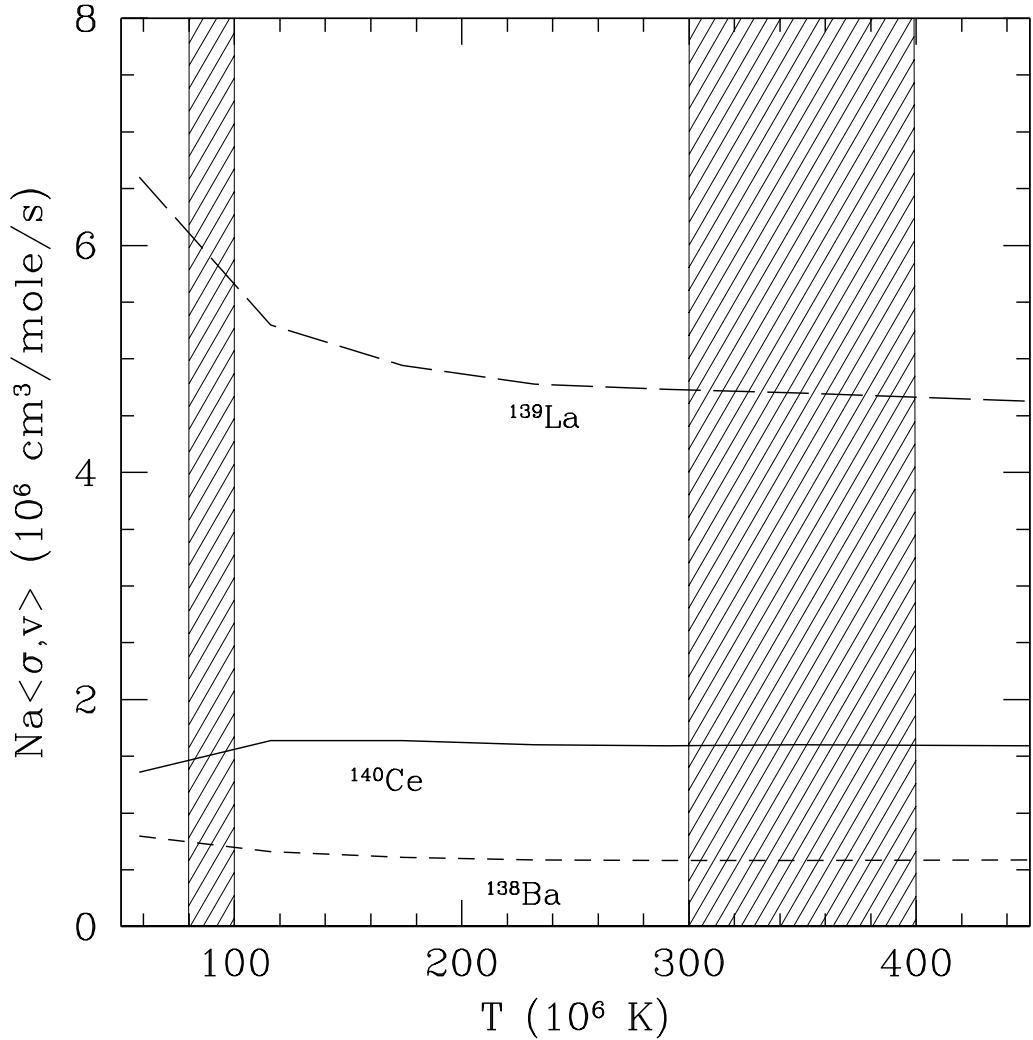}
\caption{Neutron capture reaction rates of some magic nuclei 
belonging to the heavy-s group as a function of the
temperature. The two shaded areas indicate the 
ranges of temperature experienced during the \ctan and the
\nean burnings, respectively.
Note the diversity  of $^{140}$Ce
with respect to the nearby magic nuclei.}\label{rates}
\end{figure}

The pure s-process pollution in M22 may be obtained by 
subtracting the average heavy elements composition of the stars 
belonging to the bluer sequence, those showing r-process enrichment only,
to that of the stars belonging to the redder sequence, those showing
both r and s enhancements \citep[see][]{roederer2011}. First of all,  
$\Delta t$ has been estimated by means of 5 spectroscopic indexes, namely: [Pb/Y],
[Pb/La], [Pb/Ba], [La/Y] and [Ba/Y]. Note that since La and Y are the elements with
the smaller spectroscopic errors the [La/Y] is, in principle, the best $\Delta t$
indicator. A weighted average of these five indexes 
leads to $\Delta t=144 \pm 49$ Myr. 
In Figure \ref{M22} we compare the measured s-process overabundances with those 
predicted for  $\Delta t=149$ Myr. As usual, the theoretical predictions have been shifted
to match the observed La overabundance. Such a shift mimics a certain dilution of the
gas ejected by first generation  (see equation 11).
As a whole, the agreement between the observed and the theoretical chemical pattern is
quite good. Few exceptions deserve a closer examination. 
The most evident concerns the Pb abundance, 
which is overestimated by our calculations. Lead 
is mainly produced by the radiative \ctan burning. One may argue that 
our models overestimate the extension of the $^{13}$C pockets in stars with mass
larger than 3 M$_\odot$. However,
smaller pockets cannot be the solution of this discrepancy, because in 
this case also the heavy s, from Ba to Hf, would be significantly reduced, 
thus leading to [hs/ls]$\ll 0$, in contrast with the observations. 
Nevertheless, as   
discussed in section \ref{nsources}, a reduction of the 
predicted Pb yield can be obtained by decreasing the neutrons over seeds excess.
It may be obtained in various ways: i) reducing the amount 
of $^{13}$C or ii) increasing the amount of iron seeds or
iii) increasing the amount of poisons. In a recent paper 
\citep{piersanti2013} we show that in low-metallicity stars, mixing induced by rotation, i.e., 
Goldreich-Schubert-Fricke instability and meridional 
circulation, both operating in the He-rich intershell during the interpulse periods,
increases the amount of $^{14}$N (the main poison) and iron (the main seed) 
within the $^{13}$C pocket, leaving unaltered the 
total amount of $^{13}$C. The net result is a significant reduction of the Pb (and Bi) yield, 
while the light-s are marginally enhanced and the heavy-s are practically unaffected.
A second discrepancy between our theoretical predictions and the observed compositions 
concerns Ce. In this case the theoretical expectation is smaller than the observed value.
Note that the other elements belonging to the heavy-s group, from Ba to Nd are very 
well reproduced. Then, uncertainties in the nuclear reaction rates involved in the Ce synthesis 
cannot be excluded. A check in the KADONIS database \citep{kadonis2009} discloses
that the neutron capture on $^{140}$Ce, the neutron magic isotope of Ce, 
steeply increases 
between 5 and 10 KeV, corresponding to temperatures between 50 and 120 MK, 
while it remains almost constant
for larger energies and up to 35 KeV (about 400 MK). 
This behavior substantially differs
from that of nearby magic nuclei, $^{138}$Ba and $^{139}$La,
whose neutron capture rates, 
in the same range of temperature, smoothly decrease (see Figure \ref{rates}). 
Such a peculiarity of $^{140}$Ce is likely due to 
the lack of low-energy resonances for the compound nucleus. 
As a result, the different temperature dependence of the
production channel, $^{139}$La$(n,\gamma)^{140}$La$(\beta^-)^{140}$Ce, and 
the destruction channel, $^{140}$Ce$(n,\gamma)^{141}$Ce,
favors the Ce production at low temperature ($T < 100$ MK), 
as it occurs in the case of the radiative \ctan burning. 
On the contrary, a lower Ce overabundance with respect to La 
is expected at larger temperature, 
as it occurs in the convective \nean neutron burst. Note that a 15\% reduction
of the $^{140}$Ce neutron capture cross section in the energy range 
25-35 KeV would reconcile the observed Ce overabundance with the 
corresponding theoretical prediction. A similar result could be 
obtained by increasing the $^{139}$La$(n,\gamma)^{140}$La cross 
section. These variations are probably within
the experimental uncertainties of these nuclear processes.

Other discrepancies between our predictions and M22 heavy elements composition 
regard elements whose abundance is not only determined by the AGB 
s-process nucleosynthesis, but may receive contributions from other processes.
For instance, Rh (Z=45) is usually ascribed to the r-process, while Cu (Z=29) 
and Zn (Z=30) are mainly produced by the weak-s process. 
We recall that the simplified multiple generation model here adopted 
does not include 
the possible heavy element pollution due to first-generation stars 
with M$>$6 M$_\odot$. 

In spite of the mentioned discrepancies, the comparison between the 
theoretical predictions and the M22 observed chemical pattern is 
encouraging. The derived dilution factor, as defined in equation 11, is $d=0.66$.
It implies that the material cumulatively ejected
by first-generation stars with M$\le 6$ M$_\odot$ should account 
for about 2/3 of the gas from which the second generation forms.
Finally, note that the derived delay time is smaller than the maximum age spread 
estimated from the double sub-giant branch observed in M22 \citep{marino2012}.    

\begin{figure}
\includegraphics[angle=0,width=\columnwidth]{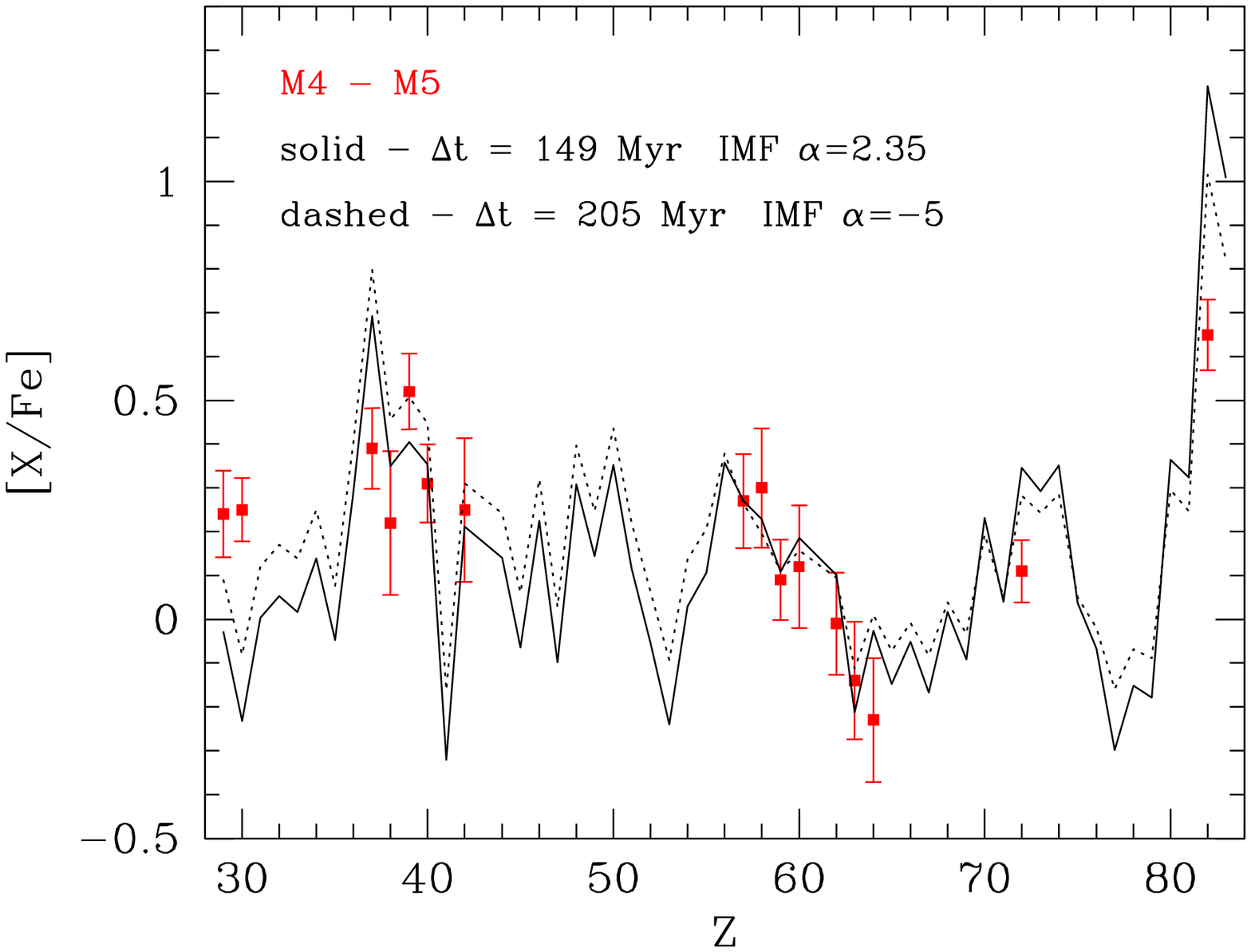}
\caption{Best fit of the average s-process chemical pattern of 
stars in M4.}\label{M4}
\end{figure}

We have repeated the same analysis with M4. In this case, owing to the lack
of s-poor cluster members,
the pure s-process component can be extracted by 
subtracting the average heavy elements composition of M5 from that of M4. 
Indeed, M5 have a metallicity very similar to that of M4, but 
it is s-process poor  \citep{yong2008}.  
The comparison with the predictions of our models is shown in Figure \ref{M4}. 
In this case we
report the best fits obtained by assuming a power law MF  
with $\alpha=2.35$ and -5. The corresponding dilution factors are $d=0.45$ 
and $d=0.67$, respectively. Note the
remarkably small value obtained in the case $\alpha=2.35$, to be compared with
the 0.66 of M22, which might be another 
evidence of the different formation history of this cluster.

Also in this case the theory overestimates the Pb abundance, even 
in the most 
favorable case of an extreme MF, while the overall reproduction of the 
ls and hs is always within the error bars. This occurrence reinforces the 
need of a 
mechanism, such as rotation, able to reduce the neutron-to-seed ratio
in the $^{13}$C pockets of intermediate-mass AGB stars. 

The Rb measurement in M4 is particularly interesting. 
We recall that the synthesis of this element depends on the two branchings at 
$^{85}$Kr  and $^{86}$Rb, so that its  
abundance represents a proof of the neutron densities of the s process 
(see section \ref{polluters}). 
In particular, a Rb excess, compared to the other light-s, is expected in 
the case of the convective \nean neutron burst.
The measured Rb abundance is comparable to the average light-s abundance. 
In particular,
it is larger than the abundances of Sr and Zr, but smaller than that of Y. 
The scatter of 
these light-s abundances is likely representative of the true observational error.
As shown in the upper panel of Figure 2 (see also Table 2),
[Rb/ls] $\leq -0.4$ is expected in the case of a negligible 
activation of the \nean reaction \citep[further details in][]{cristallo2011}. 
Therefore, even if the predicted value is larger than the observed one, 
the Rb measurements in M4 clearly indicate the operation of the convective \nean 
neutron-capture nnucleosynthesis in 
the s-process polluters. Nevertheless, the models
overestimate the observed Rb abundance.
Note that the theoretical prediction relies on rather uncertain nuclear
physics inputs affecting the $^{85}$Kr and $^{86}$Rb branchings (see section \ref{polluters}).
In particular, only theoretical evaluations are
available for the $^{86}$Rb neutron-capture cross section. From the KADONIS database
we derive that the MACS at 30 KeV is $202\pm163$ mbarn. A reduction of this cross section to 
the quoted lower limit would reduce the major Rb production channel in intermediate mass 
AGB stars.   On the other hand, for the $^{85}$Kr$_g$ neutron capture cross section, 
we have used the KADONIS prescriptions, also based on theoretical calculations, 
namely $55\pm45$ mbarn (at 30 KeV).
Recently \citet{raut2013} presented the first experimental evaluation of this
cross section. They obtain a 30 KeV MACS of $83^{+23}_{-38}$, which is higher than 
that provided by KADONIS, although compatible within the quoted errors. 
Note that the higher the  $^{85}$Kr$_g$ neutron capture cross section the
smaller is the amount of $^{85}$Kr$_g$ survived at the end of the TP and, in turn,
the smaller is the amount of $^{85}$Rb accumulated after its slow $\beta$ decay and,
later on, dredged up. Note that
 an increase of this cross section has minor effects on the overall $^{87}$Rb production, 
 because of the very low neutron-capture cross section 
on $^{86}$Kr (see section \ref{polluters}).
Therefore, further experimental investigations, as well as
other spectroscopic confirmations of Rb abundance in s-rich GC stars
are required to solve this problem.
 
Although the analysis of light elements cannot be limited to
the restricted range of stellar masses here considered, some
further considerations may be derived from the present study. 
Among the light elements fluorine deserves a major attention. Until now,
F enhancements have been found in AGB stars only or in 
stars polluted by an AGB companion, e.g., CEMPs stars \citep{jorissen1992,schuler2007, 
abia2010, abia2011, lucatello2011}. As shown in section \ref{polluters}, an anticorrelation
between the F production and the stellar mass is expected, so that the larger the $\Delta t$
the higher the fluorine pollution.
Recently, \citet{dorazi2013b}   
derived F abundance for a small sample of stars in M22, namely 3 r-only stars
plus 3 r+s stars. Although no significant variations of the [F/Fe] is found,
the small difference in the average Fe abundance of the two groups of stars
may suggest a moderate F enhancement in the r+s stars with respect to the 
r-only. However, \citet{delaverny2013} argue that these measurements are
affected by an incorrect identification of continuum fluctuations as HF signature
and a wrong correction of the stellar radial velocity. 
Owing to these uncertainties, further investigations are required before considering
fluorine measurements in the more general context of GC MP scenarios.

A further comment concerns the sodium pollution by stars with M$\le$6 M$_\odot$.
As we have reported in section 3, 
these stars release a not negligible Na yield. Such a Na is not produced by the
HBB, because in these models the 
convective envelope never attains 
the layer where the Ne-Na cycle is active. Instead, 
the Na enhancements are a consequence of 
the second and, to a less extent, of the third dredge up. If the 
duration of the pollution phase is sufficiently large, 
this contribution should be considered in addition to the  Na pollution 
possibly caused by more massive AGB, super-AGB and/or massive stars.  
This occurrence would imply a deviation from a straight O-Na anticorrelation, 
leading to a certain spread of Na in stars with similar O.
Similarly, since Mg and Al isotopes participate
to the neutron-capture nucleosynthesis, 
also the Mg-Al anticorrelation may be affected by the 
pollution of intermediate mass AGB stars. In particular,  
$^{25}$Mg is produced by the \nean reaction in the He-rich intershell 
and both Mg and Al isotopes
are produced by neutron capture chains starting from $^{22}$Ne and $^{23}$Na.

\section{Conclusions}
In this paper we presented new models of low-metallicity AGB stars with mass in the range 2-6 M$_\odot$.
The heavy elements yields of these models allow us to reproduce
most of the observed features of the s-process main and strong components,
 as shown by stars of some GC stellar populations.
The comparison between the theoretical predictions and the observed overabundance 
of s elements has been done by adopting a simple MP model for the
early GC history. This model implies two main temporal steps, namely:

1) a first stellar generation forms from pristine gas 
whose heavy element composition is that typical of the bulk of the galactic halo, 
i.e., r rich, but s poor.

2) after about $150\pm 50$ Myr, a second stellar generation forms within a newborn 
GC from the gas ejected by the stars of the first generation, 
possibly diluted with some amount of pristine gas.  

The first generation may or may not be a member of the cluster where the second
generation is observed. In other words, the pollution may be either primordial 
or internal to the cluster (self pollution). 

According to this picture, if the
star formation definitely halts in less than $\sim 50$ Myr,
namely before that intermediate-mass stars (M$\le6$ M$_\odot$) have time 
to evolve up to the AGB phase and pollute the interstellar gas, the
GC will be s-process poor. This occurrence explains
why s-process enhancements are so rare in GCs. It also implies
that the more massive stars, whose lifetime is shorter than 50 Myr,
do not substantially contribute to the main and strong components of the s process. 
On the contrary, these stars should be responsible, fully or partially,
for the more common variations of C, N, O, Na, Mg, Al and 
other light elements.   
For this reason, a more powerful and complete pollution model may be obtained by  
coupling the yields here presented to those of more massive stars. 
On the other hand, physical phenomena not yet included in
the present stellar models, such as rotation, may also improve the theoretical tool.

We are grateful to F. Kappeler and I. Dillman, for they help in interpreting the 
KADONIS reaction rates,
and to D. Yong for providing us the M4 and M5 data in a computer readable form. 
The present work has been support by the PRIN-INAF 2010 and FIRB-MIUR 2008 (RBFR08549F-002) 
programs. Extended Tables of the models presented in this paper are available in the 
FRUITY database (fruity.oa-teramo.inaf.it). 

\bibliographystyle{apj}
\bibliography{letter}


\clearpage

\begin{deluxetable}{ccccccccc}
\tablecolumns{9}
\tablewidth{0pc}
\tablecaption{Physical properties of the computed stellar models: initial mass (M$_\odot$), total lifetime (Myr),
n. of TPs followed by a TDU, total dredge-up mass ($10^{-2}$ M$_\odot$),
final core mass (M$_\odot$), total ejected mass (M$_\odot$), maximum peak temperature attained at the bottom of the 
convective shells generated by thermal pulses (MK), maximum temperature attained at 
the bottom of the convective envelope during the AGB (MK), neutron density corresponding to 
the maximum peak temperature attained at the bottom of the convective shells 
generated by thermal pulses ($10^{13}$ neutrons/cm$^3$).\label{tabphys}}
\tablehead{\colhead{Mass} & \colhead{lifetime}  & \colhead{n. TPs} & \colhead{M$_{TDU}$} & \colhead{M$_H^f$} & \colhead{ejected mass} & \colhead{$T_{csh}^{MAX}$} & \colhead{$T_{ce}^{MAX}$} & \colhead{$n_n$}}
\startdata
2.0 & 861  &  10   & 7.5   &   0.670 &  1.330 & 320  &    4 & $<0.1$  \\   
2.5 & 493  &  13   & 6.7   &   0.725 &  1.775 & 342  &    6 & 0.1  \\   
3.0 & 302  &  15   & 3.9   &   0.814 &  2.186 & 347  &   11 & 0.5  \\   
3.5 & 205  &  19   & 3.9   &   0.850 &  2.650 & 365  &   19 & 0.8  \\   
4.0 & 149  &  23   & 3.9   &   0.875 &  3.125 & 366  &   22 & 0.9  \\   
4.5 & 113  &  29   & 3.9   &   0.909 &  3.591 & 367  &   30 & 1.1  \\   
5.0 &  90  &  35   & 3.5   &   0.947 &  4.053 & 371  &   43 & 1.8  \\   
5.5 &  73  &  48   & 3.0   &   0.996 &  4.504 & 373  &   73 & 1.5  \\   
6.0 &  61  &  72   & 2.0   &   1.052 &  4.949 & 371  &   91 & 0.8  \\
\enddata
\end{deluxetable}

\clearpage

\begin{deluxetable}{cccccccccc}
\tablecolumns{10}
\tablewidth{0pc}
\rotate
\tablecaption{Average composition of the ejected material.\label{tabchim}}
\tablehead{ \colhead{Mass} & \colhead{$\left [ \frac{C+N+O}{Fe} \right ]$} & \colhead{$\left [ \frac{F}{Fe} \right ]$} & \colhead{$\left [ \frac{Na}{Fe} \right ]$} & \colhead{$\left [ \frac{Rb}{Fe} \right ]$} & \colhead{$\left [ \frac{Y}{Fe} \right ]$} & \colhead{$\left [ \frac{La}{Fe} \right ]$} & \colhead{$\left [ \frac{Pb}{Fe} \right ]$} & \colhead{$\left [ \frac{La}{Y} \right ]$} & \colhead{$\left [ \frac{Pb}{Y} \right ]$}}
\startdata
  2.0 &  1.69 &  2.02 &   0.80 &   0.53 &   0.85 &   1.37 &   2.69 &   0.52 &   1.84 \\
  2.5 &  1.57 &  1.91 &   0.75 &   0.66 &   0.76 &   1.24 &   2.60 &   0.48 &   1.84 \\
  3.0 &  1.38 &  1.57 &   0.55 &   0.98 &   1.00 &   1.36 &   2.49 &   0.36 &   1.49 \\
  3.5 &  1.32 &  1.31 &   0.68 &   0.98 &   0.83 &   1.07 &   2.20 &   0.24 &   1.13 \\
  4.0 &  1.29 &  1.06 &   0.80 &   1.06 &   0.81 &   0.87 &   1.95 &   0.06 &   1.13 \\
  4.5 &  1.24 &  0.76 &   0.87 &   1.06 &   0.76 &   0.63 &   1.60 &  -0.13 &   0.84 \\
  5.0 &  1.19 &  0.49 &   0.91 &   1.10 &   0.78 &   0.48 &   1.26 &  -0.30 &   0.48 \\
  5.5 &  1.14 &  0.28 &   0.97 &   1.07 &   0.77 &   0.44 &   1.00 &  -0.33 &   0.23 \\
  6.0 &  0.97 & -0.08 &   0.97 &   0.81 &   0.53 &   0.22 &   0.37 &  -0.31 &  -0.16 \\
\enddata
\end{deluxetable}

\clearpage

\clearpage
 
\clearpage

\clearpage

\clearpage

\clearpage

\clearpage

\clearpage

\clearpage

\clearpage

\clearpage

\clearpage

\clearpage

\clearpage

\clearpage

\end{document}